\documentclass[final]{siamltex}
\usepackage{amsmath,amssymb}
\usepackage{caption}
\usepackage{subcaption}
\usepackage{graphicx,import}
\usepackage[utf8x]{inputenc}
\usepackage{url}
\usepackage{array}
\usepackage[usenames]{color}
\usepackage{algorithm}
\usepackage{algorithmic}
\usepackage{tikz}
\usepackage{rotating}
\usepackage{siunitx}
\usepackage{relsize}
\usepackage[colorinlistoftodos,prependcaption,textsize=small]{todonotes}

\newcommand{\foot}[1]{\ensuremath{_{\text{#1}}}}
\newcommand{\head}[1]{\ensuremath{^{\text{#1}}}}
\newboolean{SHOWCOMMENTS}
\setboolean{SHOWCOMMENTS}{true}
\ifthenelse{\boolean{SHOWCOMMENTS}}{%

}

\def \scaleparam {\theta}
\def \grossemm {\mu}

\begin{document}
\title{Scheduling massively parallel multigrid 
for multilevel Monte Carlo Methods}

\author{Bj\"orn~Gmeiner\thanks{Institute of System Simulation,
		University Erlangen-Nuremberg, 
		91058 Erlangen, Germany (\tt{bjoern.gmeiner@fau.de}, \tt{ulrich.ruede@fau.de})}
\and Daniel Drzisga\thanks{Institute for Numerical Mathematics,
                 Technische Universit\"at M\"unchen,
		85748 Garching, Germany
                 ({\tt drzisga@ma.tum.de, wohlmuth@ma.tum.de}), Partly funded by WO671/11-1 (DFG)}
\and Ulrich~R\"ude\footnotemark[1] \and Robert
Scheichl\thanks{Dept. 
Mathematical Sciences, University of
  Bath, Bath BA2 7AY, UK (\tt{r.scheichl@bath.ac.uk})}
\and Barbara~Wohlmuth\footnotemark[2]}

\maketitle
%

\begin{abstract}
The computational complexity of naive, sampling-based uncertainty
quantification for 3D partial differential equations is extremely high. 
Multilevel approaches, such as multilevel Monte Carlo (MLMC), 
can reduce the complexity significantly, but to exploit them fully in
a parallel environment, sophisticated scheduling strategies are needed.
Often fast algorithms that are executed in parallel are essential to
compute fine level samples in 3D, whereas to compute individual coarse 
level samples only moderate numbers of processors can be employed efficiently.
We make use of multiple instances of a parallel multigrid solver
combined with advanced load balancing techniques. In particular, we 
optimize the concurrent execution across the three layers of the MLMC method:
parallelization across levels, across samples, and across the spatial grid. 
The overall efficiency and performance of these methods will be analyzed.
Here the ''scalability window'' of the multigrid solver is revealed as being essential,
i.e., the property that the solution can be computed with a range of
process numbers while maintaining good parallel efficiency.
We evaluate the new scheduling strategies in a series of numerical tests,
and conclude the paper demonstrating large 3D scaling experiments.
\end{abstract}
\section{Introduction}
%
Data uncertainties are ubiquitous in many application fields, such as
subsurface flow or climate prediction. 
Inherent uncertainties in input 
data propagate to uncertainties in quantities of interest,
such as the time it takes pollutants leaking from a
waste repository to reach a drink water well. This
situation has driven the development of novel
uncertainty quantification (UQ) methods; most commonly, using partial
differential equations (PDEs) to model the physical processes and
stochastic models to incorporate data uncertainties.
Simulation outputs are then statistics (mean, moments, cumulative
distribution function) of the quantities of interest. However, typical 
sampling-and-averaging techniques for computing statistics
quickly become infeasible, when each sample involves the numerical
solution of a PDE.

\subsection{Mathematical model and UQ methods}
Let us consider an abstract, possibly
nonlinear system of PDEs with uncertain data
\begin{equation}
\label{pdemodel}
\mathcal{M}(u;\omega) \ = \ 0,
\end{equation}
where the solution $u$ is sought in some suitable space $V$ of
functions $v:D \subset \mathbb{R}^d \to \mathbb{R}^k$ with $k \in
\mathbb{N}$ and $D$ open and bounded, subject to suitable boundary
conditions. $\mathcal{M}$ is a differential operator depending on a
set of random parameters parametrised by an element $\omega$ of the
abstract sample space $(\Omega,\mathcal{F},\mathbb{P})$ that
encapsulates the uncertainty in the data, with $\Omega$ the set of all
outcomes, $\mathcal{F}$ the $\sigma$-algebra (the ``set'' of all events), and $\mathbb{P}$ the associated probability measure.
As a consequence the solution $u$ itself is a random field,
i.e.~$u=u(x,\omega)$, with realizations in $V$.

We are typically only interested in functionals $Q(u) \in
\mathbb{R}$ of $u$. To compute them we
need to approximate the solution $u$ numerically, e.g.\ using finite
element methods, which introduces bias error. The cost $\mathcal{C}$
typically grows inverse proportionally to some power of the bias
error, i.e. $\mathcal{C} = \mathcal{O}(\varepsilon^{-r})$ where
$\varepsilon$ denotes the bias error tolerance. 
This is a challenging computational task that requires
novel methodology combined with cutting-edge parallel computing for two reasons: 
firstly, real life applications lead to 
PDE systems in three dimensions that often can only be solved effectively and accurately
on a parallel computer (even without data uncertainties);
secondly, typical uncertainties in applications, such
as a random diffusion coefficient $k(x,\omega)$, are spatially varying on many scales 
and cannot be described by a handful of stochastic parameters.
This limits considerably the types of UQ methods that are applicable.

For low dimensional problems, stochastic Galerkin,
stochastic collocation and polynomial chaos methods
have been shown to provide efficient and powerful UQ tools
(see, e.g., \cite{ghanem1991stochastic,XiuKarniadakis,Lord_etal} and
the references therein), but in general their complexity grows
exponentially with the stochastic dimension. The cost of sampling
methods, such as, e.g., Monte Carlo,
does not grow with the stochastic dimension, 
but classical Monte Carlo is notoriously slow to converge. 
Multilevel Monte Carlo (MLMC) simulation
\cite{giles,cliffe2011multilevel} can help to significantly accelerate the convergence.
It has been applied and extended to a range of 
applications, see \cite{barth,mishra,collier,elfverson,hoang}. 

The idea of MLMC is to reduce algorithmic complexity
by performing as much computational work as possible
on coarse meshes. 
To this end, MLMC uses a hierarchy of 
discretisations of \eqref{pdemodel} of increasing accuracy 
to estimate statistics of $Q(u)$ more efficiently, 
i.e.\ using a large number of coarse samples to fully capture the variability,
but only a handful of fine samples to eliminate the bias due
to the spatial discretisation. Here, we employ multilevel methods
not only to accelerate the stochastic part, 
but also to provide a scalable solver for individual realizations of \eqref{pdemodel}. 

\subsection{Parallel methods and algorithms}
Current leading-edge supercomputers provide a peak performance
in the order of a hundred petaflop/s
(i.e.~$10^{17}$ floating point operations per second) \cite{strohmaier2015top500}.
However, all these computers draw their computational power from 
parallelism, with current processor numbers already
at $P\foot{max}\approx10^7$ see \cite{Dongarra2016}.
The technological trend indicates that future exascale computers may  
use $P\foot{max}\approx10^9$.
Consequently, designing efficient fast parallel algorithms 
for high performance computers is a challenging task
today and will be even more so in the future.

MLMC methods are characterized by 
three algorithmic levels that are potential candidates for parallel execution.
As in a standard Monte Carlo method, the algorithm
uses a sequence of classical deterministic problems (samples)
that can be computed in parallel.
The size of these subproblems
varies depending on which level of resolution
the samples are computed on.
We therefore distinguish between parallelism within an MLMC level 
and parallelism across MLMC levels. 
The third algorithmic level is the solver for each deterministic PDE
problem which can be parallelized itself. 
Indeed, the total number of
samples on finer MLMC levels is typically moderate, so that the first
two levels of parallelism will not suffice to exploit $P\foot{max}$ 
processors.
Parallel solvers for elliptic PDEs are now able to solve systems with 
\num{1.1e13} degrees of freedom on petascale machines \cite{perf-analysis-16}
with compute times of a few minutes using highly parallel multigrid methods
\cite{chow2006survey,gmeiner2014parallel}.
In this paper, we will illustrate for a simple model problem
in three spatial dimensions,
how these different levels of parallelism can be combined
and how efficient parallel MLMC strategies can be designed. 

To achieve this, we extend the
massively parallel Hierarchical Hybrid Grids (HHG) framework \cite{bergen2004hierarchical,gmeiner2015towards}
that exhibits excellent strong and weak scaling behavior
\cite{hager2010introduction,baker2012scaling}
to the MLMC setting. We use the fast multigrid solver in HHG to
generate spatially correlated samples of
the random diffusion coefficient, as well as to solve the resulting
subsurface flow problems efficiently. Furthermore, the hierarchy of 
discretisations in HHG provides the ideal multilevel framework for the 
MLMC algorithm.

Parallel solvers may not yield linear speedup and the efficiency may
deteriorate on a large parallel computer system when the problems 
become too small. In this case, too little work can be executed 
concurrently and the scalar overhead dominates.
This effect is well-known and can be understood prototypically in the form of 
Amdahl's law \cite{hager2010introduction}.
In the MLMC context, problems
of drastically different size must be solved. 
In general, a solver, when applied to a problem of given size,
will be characterized by its \emph{scalability window},
i.e., the processor range for which the parallel efficiency remains above an
acceptable threshold.
Because of memory constraints, the scalability window
will open at a certain minimal processor number.
For larger processor numbers the parallel efficiency will deteriorate 
until the scalability window closes.
In practice, additional restrictions imposed by the system and the software
permit only specific processor numbers within the scalability window
to be used.

MLMC typically leads to a large number of small problems, 
a small number of very large problems, and a fair number of
intermediate size problems. On the coarser levels, the problem 
size is in general too small to use the full machine. The problem is outside 
the scalability window and solver-parallelism alone is insufficient. 
On the other hand, the efficiency of parallelization across samples
and across levels typically does not deteriorate, since only little data must
be extracted from each sample to compute the final result of the UQ
problem. However, on finer levels we may not have 
enough samples to fill the entire machine.
Especially for adaptive MLMC,
where the number of samples on each level is not known a priori but 
must be computed adaptively using data from all levels, 
this creates a challenging load balancing problem.

A large scalability window of the solver is essential 
to devise highly efficient execution strategies, but finding 
the optimal schedule is restricted by a
complex combination of mathematical and technical constraints.
Thus the scheduling problem becomes
in itself a high-dimensional, multi-constrained,
discrete optimisation problem.
Developing suitable approaches in this 
setting is one of the main objectives of this paper. See 
\cite{vsukys2014adaptive,vsukys2012static} for earlier static and
dynamic load balancing approaches.

The paper is structured as follows: 
In Section \ref{sec:mlmc}, we briefly review the MLMC method and its 
adaptive version. Section \ref{sec:sampling} introduces the model
problem. Here, we use an alternative PDE-based sampling technique for
Mat\'ern covariances
\cite{lindgren2011explicit,PF10} that allows us to reuse the parallel 
multigrid solver. In Sections \ref{sec:class} and \ref{sec:impl}, 
we define a classification of different parallel execution strategies
and develop them into different parallel scheduling approaches. 
In Section \ref{sec:sched},
we study the parallel efficiency of the proposed strategies and 
demonstrate their flexibility and robustness,
before finishing in
Section \ref{sec:error} with 
large-scale experiments on advanced supercomputer systems.

\section{The Multilevel Monte Carlo method}
%
\label{sec:mlmc} 
To describe the MLMC method, 
we assume that we have a hierarchy of finite element (FE)
discretisations of \eqref{pdemodel}.
Let $\{V_\ell\}_{\ell\ge 0}$ be a nested sequence of FE spaces with
$V_\ell \subset V$, mesh size $h_\ell > 0$ and $M_\ell$ degrees of
freedom. In the Hierarchical Hybrid Grids (HHG) framework
\cite{bergen2004hierarchical,gmeiner2015towards}, the underlying 
sequence of FE meshes is obtained via uniform mesh refinement from 
a coarsest grid $\mathcal{T}_0$, and thus $h_\ell
\simeq 2^{-\ell}h_0$ and $M_\ell \simeq 2^{3 \ell} M_0$ in three space 
dimensions. 

Denoting by $u_\ell = u_\ell(x, \omega) \in V_\ell$ the FE
approximation of $u$ on Level $\ell$, we have
\begin{equation}
\label{pdemodel_l}
\mathcal{M}_\ell(u_\ell;\omega) = 0, \qquad \ell \ge 0.
\end{equation}
Here, the (non)linear operator $\mathcal{M}_\ell$ and the functional
of interest
$Q_\ell(u_\ell,\omega)$ may also 
involve numerical approximations.

\subsection{Standard Monte-Carlo Simulation}
\label{sec:singlelevel}
The standard Monte Carlo (MC) estimator for the expected value
$\mathbb{E}[Q]$ of $Q(u)$ on level $L \ge 0$ is given by
\begin{equation}\label{eq:MC_estimator}
\widehat{Q}^{\text{MC},N}_{L} = \frac{1}{N} \sum_{i=1}^N 
Q_L^{i}\,,
\end{equation}
where $Q_L^{i} = Q_L(u_L^{i},\omega^{i})$, $i=1,\ldots,N$, are $N$
independent samples of $Q_L(u_L)$.

There are two sources of error: (i) The {\em bias error} due to the FE
approximation. Assuming that $|Q_L^{i} - Q(u^{i},\omega^{i})| = 
\mathcal{O}(M_L^{-\alpha})$, for almost all $\omega^i$ and a
constant $\alpha>0$, it follows directly that there exists a
constant $C_b$, independent of $M_L$, such that
\begin{equation}\label{eq:M_alpha}
|\mathbb{E}[Q_L-Q]| \le C_b M_L^{-\alpha} \leq \varepsilon_b
\end{equation} 
 for $M_L \ge
(\varepsilon_b/C_b)^{1/\alpha}$ (cf.~\cite{TSGU:2012}).

(ii) There is a {\em sampling error}
due to the finite number $N$ of samples in \eqref{eq:MC_estimator}.

The total error is typically quantified via the \textit{mean square error} (MSE), given by
\begin{equation}
\label{eq:MC_MSE}
e\left(\widehat{Q}^{\text{MC},N}_L\right)^2 := \mathbb{E}[(\widehat{Q}^{\text{MC},N}_L - \mathbb{E}[Q])^2] = \left(\mathbb{E}[Q_L
- Q]\right)^2 + N^{-1}\mathbb{V}[Q_L],
\end{equation}
where $\mathbb{V}[Q_L]$ denotes the variance of the random variable
$Q_L(u_L)$. The first term in
\eqref{eq:MC_MSE} can be bounded in terms of \eqref{eq:M_alpha}, and
the second term in  is smaller than a sample
tolerance $\varepsilon_s^2$ if $N \ge
\mathbb{V}[Q_L] \varepsilon_s^{-2}$. We note that for $L$ 
sufficiently large, $\mathbb{V}[Q_L]
\approx \mathbb{V}[Q]$. To ensure that the total MSE is less than $\varepsilon^2$ we choose 
\begin{equation}
\label{def:eps}
\varepsilon_s^2 = \theta \varepsilon^2 \quad \text{and} \quad \varepsilon_b^2 = (1-\theta)\varepsilon^2, \quad \text{for any fixed} \ \  0 < \theta < 1.
\end{equation}

Thus, to reduce \eqref{eq:MC_MSE} we need to choose a sufficiently fine FE mesh and a sufficiently large number of samples. This very
quickly leads to an intractable problem for complex PDE problems in 3D.
The cost for one sample
$Q_L^{i}$ of $Q_L$ depends on the complexity of the FE solver and of
the random field generator. Typically it will grow
like $C_c M_L^\gamma$, for some $\gamma \ge 1$ and some constant $C_c$,
independent of $i$ and of $M_L$. Thus, the total cost to achieve a MSE
$e(\widehat{Q}_L^{\text{MC},N})^2 \le \varepsilon^2$ (the {\em $\varepsilon$-cost}) is
\begin{equation}
\text{Cost}\left(\widehat{Q}_L^{\text{MC},N}\right) = \mathcal{O}(M^\gamma N) = \mathcal{O}(\varepsilon^{-2-\gamma/\alpha}).
\end{equation}

 For the coefficient
field and for the output functional studied below, we have  only $\alpha = 1/6$.
In that case, even if $\gamma=1$, to reduce the error by a factor 2 the cost grows by a factor of $2^8 = 256$, which quickly leads to an intractable problem even in a massively parallel environment.

\subsection{Multilevel Monte-Carlo Simulation}
Multilevel Monte Carlo (MLMC) simulation 
\cite{giles,cliffe2011multilevel,barth} 
seeks to reduce the variance
of the estimator  and thus to reduce computational time,
by recursively using coarser FE models as control variates.
By exploiting the linearity of the expectation operator,
we avoid estimating $\mathbb{E}[Q]$ directly on the finest level~$L$
and do not compute all samples to the desired accuracy (bias error). 
Instead, using the
simple identity $
\mathbb{E}[Q_L] = \mathbb{E}[Q_0] + \sum_{\ell=1}^L\mathbb{E}[Y_\ell] $,
we estimate the mean on the coarsest level (Level~$0$) and correct
this mean successively by adding estimates of the expected values of 
$Y_\ell(\omega) :=
Q_\ell(u_\ell,\omega) - Q_{\ell-1}(u_{\ell-1},\omega)$, for $\ell \ge 1$. 
Setting $Y_0 := Q_0$, the MLMC estimator is then defined as
\begin{equation}\label{eq:MLMC_defn}
\widehat{Q}_L^{\text{ML}} := \sum_{\ell=0}^{L}\widehat{Y}^{\text{MC},N_\ell}_\ell \,,
\end{equation}
where the numbers of samples $N_\ell$, $\ell=0,\ldots,L$, are chosen to
minimize the total cost of this estimator for a given prescribed
sampling error (see Eqn.~\eqref{optimal_Nl} below). Note that we
require the FE solutions $u_\ell(x,\omega^{i})$ and $u_{\ell-1}(x,\omega^{i})$
on two levels to compute a sample $Y^{i}_\ell$ of $Y_\ell$, for
$\ell \ge 1$, and thus two PDE solves, but crucially both 
with the same $\omega^{i}$ and thus with the same PDE 
coefficient (see Algorithm \ref{mlmc_algorithm}).
\begin{algorithm} 
\caption{Multilevel Monte Carlo. \label{mlmc_algorithm}}
\tt
\begin{tabbing}
1. \= For all levels $\ell=0,\ldots,L$ do\\[0.5ex]
\> a. \= For $i=1,\ldots,N_\ell$ do\\
\> \> i. \ Set up \eqref{pdemodel_l} for $\omega^{i}$ on Level $\ell$ and $\ell-1$ (if $\ell > 0$).\\
\> \> ii. Compute $u_\ell(\omega^{i})$ and $u_{\ell-1}(\omega^{i})$ (if $\ell > 0$), as well as $Y^{i}_\ell$.\\[0.5ex]
\> b. \> Compute $\widehat{Y}^{\mathrm{MC},N_\ell}_{\ell} = \frac{1}{N_\ell} \sum_{i=1}^{N_\ell} Y^{i}_\ell$.\\[0.5ex]
2. \> Compute $\widehat{Q}_L^{\text{ML}}$ using \eqref{eq:MLMC_defn}.
\end{tabbing}
\end{algorithm}

The cost of this estimator is
\begin{equation}
\label{MLMC_cost}
\text{Cost}(\widehat{Q}_{L}^\text{ML}) =  \sum_{\ell=0}^{L} N_\ell
\mathcal{C}_\ell\,,
\end{equation}
where $\mathcal{C}_\ell$ is the cost to compute one sample of $Y_\ell$ on level $\ell$. 
For simplicity, we use independent samples across all
levels, so that the $L+1$ standard MC estimators in \eqref{eq:MLMC_defn} are
independent. Then, the MSE of
$\widehat{Q}_L^{\text{ML}}$ simply expands to
\begin{equation}
e\left(\widehat{Q}_L^{\text{ML}}\right)^2 = \big(\mathbb{E}[Q_L-Q]\big)^2 \; + \;
\sum_{l=0}^L N_\ell^{-1}\mathbb{V}[Y_\ell]\,.
\end{equation}
This leads to a hugely reduced variance of the estimator since both FE approximations $Q_\ell$ and $Q_{\ell-1}$
converge to $Q$ and thus $\mathbb{V}[Y_\ell]\to 0$, as $M_{\ell-1} \to \infty$.

By choosing $M_L \ge
(\varepsilon_b/C_b)^{-1/\alpha}$, we can ensure again that the bias error is
less than $\varepsilon_b$, but we still
have some freedom to choose the numbers of samples
$N_\ell$ on each of the levels, and thus to ensure that the sampling
error is less than $\varepsilon_s^2$. We will use this freedom to
minimize the cost $\text{Cost}(\widehat{Q}_{L}^\text{ML})$ in
\eqref{MLMC_cost} subject to the constraint
$\sum_{\ell=0}^L N_\ell^{-1}\mathbb{V}[Y_\ell] = \varepsilon_s^2$, 
a simple discrete, constrained optimization problem with
respect to $N_0, \ldots, N_L$ (cf. \cite{giles,cliffe2011multilevel}). It leads to
\begin{equation}
\label{optimal_Nl}
N_\ell \ = \ \varepsilon_s^{-2} \left(\sum_{\ell=0}^L \sqrt{\mathbb{V}[Y_\ell]
    \mathcal{C}_\ell} \right) \, \sqrt{\frac{\mathbb{V}[Y_\ell]}{C_\ell}}.
\end{equation}

Finally, under the assumptions that 
\begin{equation}
\label{def:var}
\mathcal{C}_\ell \; \le \; C_c M_\ell^\gamma \quad \text{and} \quad \mathbb{V}[Y_\ell] \le C_v \, M_\ell^{-\beta}\,,
\end{equation}
for some $0< \beta \le 2\alpha$ and $\gamma \ge 1$ and for two
constants $C_c$ and $C_v$, independent of $i$ and of $M_\ell$, 
the $\varepsilon$-cost to achieve $e(\widehat{Q}_L^{\text{ML}})^2 \le
\varepsilon^2$ can be bounded by
\begin{equation}
\text{Cost}(\widehat{Q}_{L}^\text{ML}) \; = \; \varepsilon_s^{-2} \left(\sum_{\ell=0}^L \sqrt{\mathbb{V}[Y_\ell]
    \mathcal{C}_\ell} \right)^2 \; \le \; C_{\text{ML}} \,
\varepsilon^{-2-\max\left(0,\frac{\gamma-\beta}{\alpha}\right)} \,.
\end{equation}
Typically $\beta \approx 2 \alpha$ for smooth functionals
$Q(\cdot)$. For CDFs we typically have $\beta=\alpha$.\pagebreak

There are three regimes: $\gamma < \beta$, $\gamma=\beta$ and $\gamma >
\beta$.
In the case of the exponential covariance, typically  $\gamma > \beta$ and $\beta = 2 \alpha$ and thus $\text{Cost}(\widehat{Q}_{L}^\text{ML})  =  \mathcal{O}(\varepsilon^{-\gamma/\alpha})$, which is a full two orders of magnitude faster than the standard MC method. Moreover, MLMC is {\em optimal} for this problem, in the sense that its cost is asymptotically of the same order as the cost of computing a single sample to the same tolerance $\varepsilon$.

\subsection{Adaptive Multilevel Monte Carlo}
\label{sec:adaptive}
In Algorithm \ref{adaptive_mlmc_algorithm} we present a simple
sequential, adaptive algorithm from
\cite{giles,cliffe2011multilevel} 
that uses the computed samples to
estimate bias and sampling error and 
thus chooses the optimal values
for $L$ and $N_\ell$. 
Alternative adaptive algorithms are described in
\cite{giles_waterhouse,collier,elfverson}.  For the remainder of
the paper we will restrict to uniform mesh refinement, i.e. $h_\ell =
2^{-\ell} h_0$ and $M_\ell = \mathcal{O}(8^\ell M_0)$ in 3D. 
\begin{algorithm}
\caption{Adaptive Multilevel Monte Carlo. \label{adaptive_mlmc_algorithm}}
\tt
\begin{tabbing}
1. \= Set $\varepsilon$, $\theta$, $L=1$ and $N_0 = N_1 =  N_{\text{Init}}$.\\[0.5ex]
2. \> For all levels $\ell=0,\ldots,L$ do\\
\> a. \= Compute new samples of $Y_\ell$ until there are $N_\ell$.\\
\> b. \> Compute $\widehat{Y}^{\mathrm{MC},N_\ell}_{\ell}$ and $s^2_\ell$, and estimate $\mathcal{C}_\ell$.\\[0.5ex]
3. \> Update the estimates for $N_\ell$ using \eqref{quasioptimal_Nl}
and\\
\> if $\widehat{Y}^{\mathrm{MC},N_L}_{L} > (8^{\alpha}-1)\varepsilon_b$, increase $L \to L+1$ and set $N_L =  N_{\text{Init}}$.\\[0.5ex]
4. \> If all $N_\ell$ and $L$ are unchanged,\\
\> \> Go to 5.\\
\> Else Return to 2.\\[1ex]
5. \> Set $\widehat{Q}_{L}^\mathrm{ML} = \sum_{\ell=0}^L \widehat{Y}^{\mathrm{MC},N_\ell}_{\ell}$.
\end{tabbing}
\end{algorithm}

To estimate the bias error, let us assume that $M_\ell$ is
sufficiently large, so that we are in the asymptotic regime,
i.e. $|\mathbb{E}[Q_{\ell}-Q]| \approx C_b\, M_\ell^{-\alpha}$ in
\eqref{eq:M_alpha}. Then (cf.~\cite{elfverson}) 
\begin{equation}
\label{bias_estimate}
|\mathbb{E}[Q_{\ell}-Q]| \le \frac{1}{8^{\alpha}-1} \widehat{Y}^{\text{MC},N_\ell}_{\ell}\,.
\end{equation}
Also, using the sample estimator
$s^2_\ell := \frac{1}{N_\ell} \sum_{i=1}^{N_\ell}
\big(Y_\ell^{i}-\widehat{Y}^{\text{MC},N_\ell}_{\ell}\big)^2$ 
to estimate $\mathbb{V}[Y_\ell]$
and the CPU times from the
runs up-to-date to estimate $\mathcal{C}_\ell$,
we can estimate
\begin{equation}
\label{quasioptimal_Nl}
N_\ell \ \approx \ \varepsilon_s^{-2} \left(\sum_{\ell=0}^L \sqrt{s^2_\ell
    \mathcal{C}_\ell} \right) \, \sqrt{\frac{s^2_\ell}{C_\ell}}.
\end{equation}

\section{Model problem and deterministic solver}
%
\label{sec:sampling}
As an example, we consider an elliptic PDE in weak form: Find $u(\cdot,\omega) \in V := H^1_0(D)$ such that
\begin{equation}
\label{modweak}
\int_D \nabla v(x) \cdot \big(k(x, \omega) \nabla u(x, \omega)\big) \, \text{d}x =  \int_D f(x) v(x) \, \text{d}x, \quad \text{for all} \ v \in V \  \text{and} \ \omega \in \Omega.
\end{equation}
This problem is motivated from subsurface flow. The solution $u$ and the coefficient $k$ are random fields on $D \times \Omega$ related to fluid pressure and rock permeability. For simplicity, 
we only consider $D=(0,1)^3$, homogeneous Dirichlet conditions and a deterministic source term $f$. If $k(\cdot,\omega)$ is continuous (as a function of $x$) and $k_{\min}(\omega) := \min_{x \in \overline{D}} k(x,\omega) > 0$ almost surely (a.s.) in $\omega \in \Omega$, then it follows from the Lax-Milgram Lemma that this problem has a unique solution (cf.~\cite{CST:2011}). 
As quantities of interest in Section \ref{sec:error}, we consider
$Q(u) := u(x^*)$, for some $x^* \in D$, or alternatively  $Q(u) :=
\frac{1}{|\Gamma|} \int_{\Gamma} -k \frac{\partial u}{\partial n}\,
\text{d}s$, for some two-dimensional manifold $\Gamma \subset
\overline{D}$ can be of interest.


\subsection{Discretisation}
\label{sec:discrete}

To discretise \eqref{modweak}, for each $\omega \in \Omega$, we use
standard $\mathbb{P}_1$ finite elements on a sequence of uniformly
refined simplicial meshes $\{\mathcal{T}_\ell\}_{\ell \ge 0}$. Let
$V_\ell$ be the FE space associated with $\mathcal{T}_\ell$,
$\mathcal{N}_\ell$ the set of interior vertices, $h_\ell$ the mesh
size and $M_\ell = |\mathcal{N}_\ell|$ the number of degrees of
freedom. Now, problem \eqref{modweak} is discretised by restricting it
to functions $u_\ell, v_\ell \in V_\ell$. Using the nodal basis
$\{\phi_j : x_j \in \mathcal{N}_\ell\}$ of $V_\ell$ and expanding
$u_\ell(\cdot,\omega) := \sum_{j\in \mathcal{N}_\ell}
U^{(\ell)}_j(\omega) \phi_j$, this can be written as a linear equation
system where the entries of the system matrix  are assembled
elementwise based on on a four node quadrature formula
\begin{align*}
A^{(\ell)}(\omega) \mathbf{U}^{(\ell)}(\omega) \, = \, \mathbf{F}^{(\ell)}, \quad &\text{where}\\[1ex]
A^{(\ell)}_{i,j}(\omega) := \sum_{\tau \in \mathcal{T}_\ell} \nabla \phi_i \cdot \nabla \phi_j \big|_\tau  \, \frac{|\tau|}{4} \,\bigg( \sum_{k =1}^4 k(x_k^\tau, \omega)\bigg), \quad &\text{and} \quad  \mathbf{F}^{(\ell)}_i := \int_D f \phi_i \, \text{d}x.
\end{align*}
Here $x_k^\tau$, $1 \leq k \leq 4$ denote the four vertices of the element $\tau$.

The quantity of interest $Q(u)$ is simply approximated by $Q(u_\ell)$.
For $Q(u_\ell)$ to converge to $Q(u)$, 
as $\ell \to \infty$,  we need stronger assumptions on the random field $k$. Let $k(\cdot,\omega) \in C^{0,t}(\overline{D})$, i.e. H\"older-continuous with coefficient $t \in (0,1)$, and suppose 
$k_{\min}(\omega)$ and $\|k(\cdot,\omega)\|_{C^{0,t}}$ have bounded second moments. It was shown in \cite{TSGU:2012} that
\begin{equation}
\mathbb{E}\left[(Q(u) - Q(u_\ell))^q\right] = \mathcal{O}\left(h_\ell^{tq}\right) = \mathcal{O}\left(M_\ell^{-tq/3}\right), \quad q=1,2,
\end{equation}
Hence, the bound in \eqref{eq:M_alpha} holds with $\alpha = \frac{t}{3}$, and since 
$$
\mathbb{V}\left[Q(u_\ell) - Q(u_{\ell-1})\right] \le \mathbb{E}\left[(Q(u_\ell) - Q(u_{\ell-1}))^2\right] \le 2 \sum_{r=\ell,\ell-1} \mathbb{E}\left[(Q(u) - Q(u_r))^2\right]
$$
the bound in \eqref{def:var} holds with $\beta = 2\alpha = \frac{2t}{3}$.

%
%

\subsection{PDE-based sampling for lognormal random fields}
\label{sec:random-fields}
A coefficient function $k$ of particular interest is the lognormal random field $k(\cdot,\omega) := \exp(Z(\cdot,\omega))$, where $Z(\cdot,\omega)$ is a mean-free, stationary Gaussian random field with exponential covariance
\begin{equation}
\label{def:expcov}
\mathbb{E}[Z(x,\omega)Z(y,\omega)] = \sigma^2 \exp\left(-\frac{|x-y|}{\lambda} \right).
\end{equation}
The two parameters in this model are the {\em variance} $\sigma^2$ and the {\em correlation length} $\lambda$. Individual samples $k(\cdot,\omega)$ of this random field are in $C^{0,t}(\mathbb{R}^3)$, for any $t < 1/2$. In particular, this means that the convergence rates in \eqref{eq:M_alpha} and \eqref{def:var} are $\alpha = 1/3 - \delta$ and $\beta = 2/3 - \delta$ in this case, for any $\delta > 0$. The field $Z(\cdot,\omega)$ belongs to the larger class of Mat\'ern covariances \cite{lindgren2011explicit, Lord_etal}, which also includes smoother, stationary lognormal fields, but we will only consider the exponential covariance in this paper.

Two of the most common approaches to realise the random field $Z$ above are Karhunen-Loeve (KL) expansion~\cite{ghanem1991stochastic} and circulant embedding~\cite{DietrichNewsam,graham2011quasi}.
While the KL expansion is very convenient for analysis 
and essential for polynomial expansion methods such
as stochastic collocation, it can quickly dominate all 
the computational cost for short correlation lengths $\lambda$ in three dimensions.
Circulant embedding, on the other hand, relies on the Fast Fourier Transform, which may pose limits to scalability in a massively parallel environment.
An alternative way to sample $Z(x,\omega)$ is to exploit
the fact that in three dimensions, mean-free Gaussian fields with exponential covariance are solutions to the stochastic partial differential equation (SPDE)
\begin{equation}
\label{spde}
 (\kappa^2 - \Delta) Z(x, \omega) =^d W(x,\omega),
\end{equation}
where the right hand side $W$ is Gaussian white noise with unit variance and $=^d$ denotes equality in distribution. As shown by Whittle \cite{Whittle63}, a solution of this SPDE will be Gaussian with exponential covariance $\sigma^2 = (8 \pi \kappa)^{-1}$ and $\lambda = 2/\kappa$. 

In \cite{lindgren2011explicit}, the authors show how this SPDE can be solved using a FE discretisation and this will be the approach we use to bring our fast parallel multigrid methods to bear again. 
Since we only require samples of $k(\cdot,\omega) = \exp(Z(\cdot,\omega))$ at the vertices of $\mathcal{T}_\ell$, we discretise \eqref{spde} using again standard $\mathbb{P}_1$ finite elements.
If now $Z'_\ell(\cdot,\omega) \in V'_\ell$ denotes the FE
approximation to $Z'$, then we approximate $k(x_j, \omega)$ in our
qudrature formula 
 by $\exp(Z_\ell'(x_j, \omega))$, for all $x_j \in
 \mathcal{N}_\ell$. It was shown in
 \cite{lindgren2011explicit,simpsonetal} that $Z'_\ell$ converges in a
 certain weak sense to $Z'$ with $\mathcal{O}(M_\ell^{1/3-\delta})$,
 for any $\delta > 0$. Since \eqref{spde} is in principle posed on all
 of $\mathbb{R}^3$ we embed the problem into the larger domain 
 $\widetilde{D} := (-1,2)^3 \supset D$ with  artifical,  homogeneous 
Neumann boundary conditions on $\partial
 \widetilde{D}$ (see \cite{simpsonetal}).
%

\section{Performance parameters and  execution strategies} \label{sec:class}
%
Although MLMC methods 
can achieve better computational complexity
than standard MC methods, efficient parallel execution strategies 
are challenging and depend strongly on the performance characteristics
of the solver, in our case a multigrid method.
The ultimate goal is
to distribute the $P\foot{max}$ processors to the
different subtasks such that the total run time of the MLMC is minimal. 
This can be formulated as a high dimensional,
multi-constraint discrete optimization problem.
More precisely, this scheduling problem is in general NP-complete, see, e.g.,  
\cite{Dro09,GJ90,LKB77,Ull75} and the references therein, precluding
 exact solutions in practically relevant situations.

\subsection{Characteristic  performance parameters} \label{sec:parameters}
To design an efficient scheduling strategy, we rely on 
predictions of the
time-to solution and have to take into account fluctuations.
Static strategies thus
possibly suffer from a significant load imbalance and may
result in poor parallel efficiency. Dynamic strategies, such as the
greedy load balancing algorithms in~\cite{vsukys2014adaptive}, which take
into account run-time data are more robust, especially when run-times
vary strongly within a level.

For the best performance, the number of processors $P_\ell$ 
per sample on level $\ell$ 
should lie within the scalability window
$\{  P_\ell^{\min},  P_\ell^{\min}+1,\ldots, P_\ell^{\max}\}$ of the PDE solver,
where the parallel
efficiency is above a prescribed threshold of, e.g., 80\%.
Due to machine constraints, $P_\ell$ may be restricted to a subset,
such as
 $\{  P_\ell^{\min},  2 P_\ell^{\min}, \ldots, 2^S  P_\ell^{\min}\}$,
where $S\in \mathbb{N}_0$ characterizes the size of the scalability window and $P_\ell^{\max} = 2^S P_\ell^{\min}$. Efficient implementations
of 3D multigrid schemes such as, e.g., within HHG 
\cite{bergen2004hierarchical,gmeiner2015performance,gmeiner2015towards},
have excellent strong scalability and
a fairly large scalability window, with a typical value of
$S=4$ for a parallel efficiency threshold of $80\%$. The  HHG solver has not only
good strong scaling properties but also exhibits excellent weak scalability.
We can thus assume that $P_\ell^{\min} = 2^{3\ell} P_0^{\min} $ and $P_\ell^{\max} = 2^{3\ell} P_0^{\max} $ for PDEs in 3D. The value of 
$P_0^{\min}$ is the number of processors for which the 
main memory capacity is fully utilized. Multigrid PDE solvers
typically achieve the best parallel efficiency for $P= P_\ell^{\min}$,
when each subdomain is
as large as possible and the ratio of computation to communication is
maximal (cf.~\cite{gmeiner2014parallel}). 

\begin{figure}[ht]
\resizebox{0.47\textwidth}{!}{\relsize{1.7} 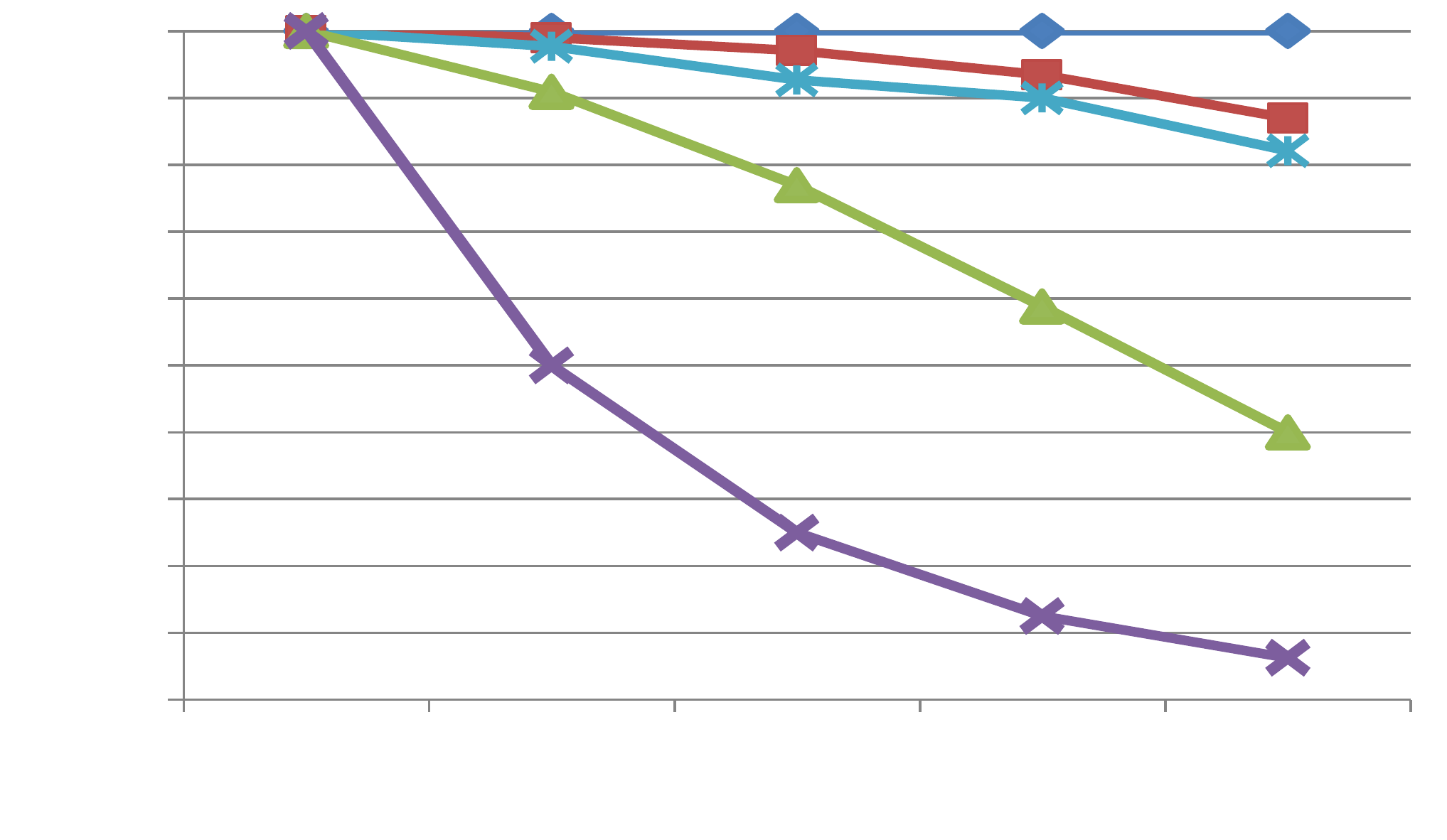}
\hfill
\resizebox{0.5\textwidth}{!}{\relsize{1.7} 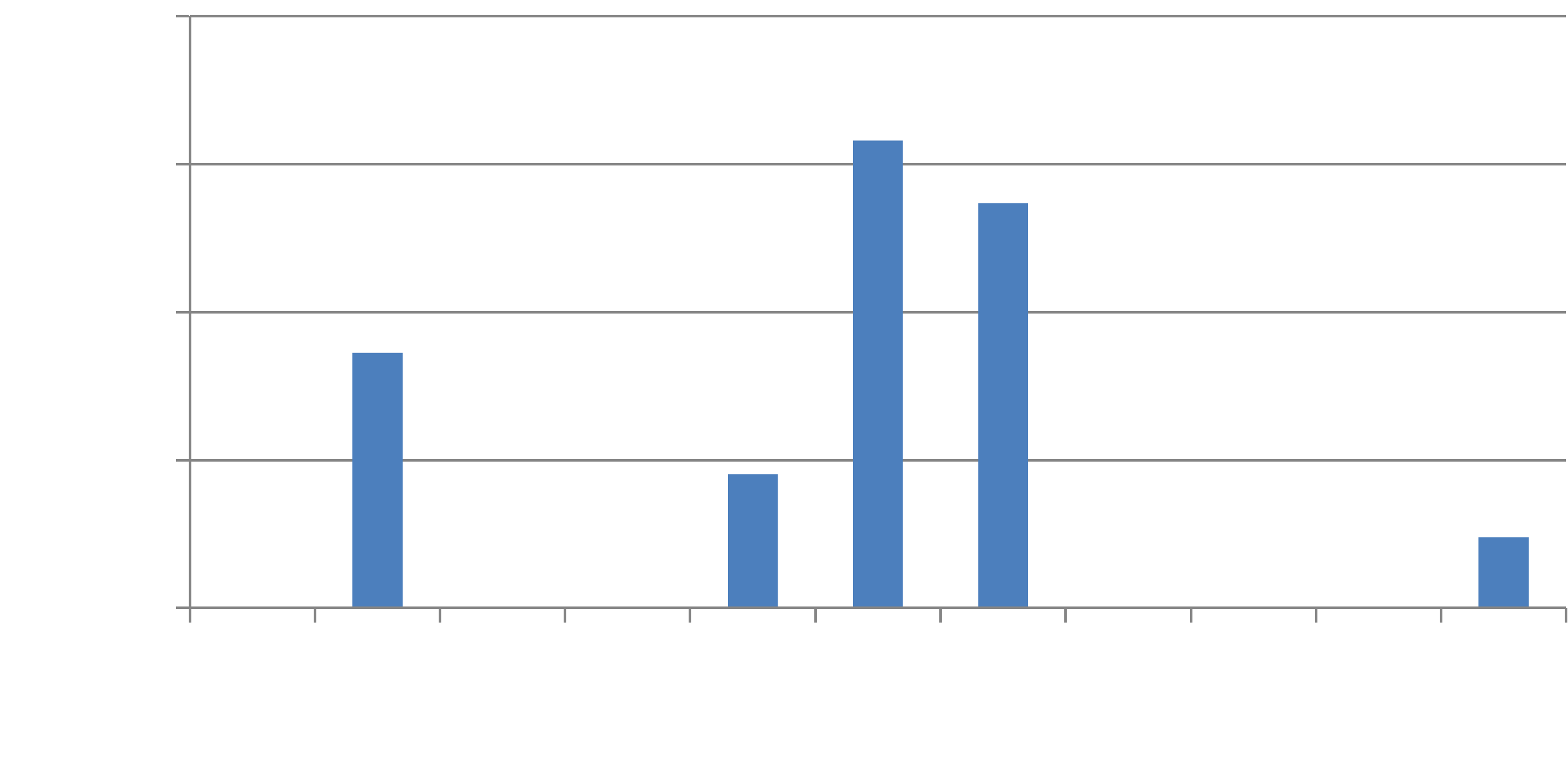}
\caption{\label{fig:strongam} Left: Parallel efficiency for different serial fraction parameters $B$, Right: Example of a run-time histogram for a multigrid solver  using full multigrid-cycles.}
\end{figure}

In the following, the time-to solution for the $i$th sample on level
$\ell$ executing on $2^\scaleparam P_\ell^{\min} =
2^{3\ell+\scaleparam} P_0^{\min}$ processors is
denoted by $t(i,\ell,\scaleparam)$. We
assume that 
\begin{equation} \label{eq:time-for-sample}
t (i,\ell,\scaleparam) \approx C_{\ell,\scaleparam}(\omega^{i}) \, t_{\ell, \scaleparam}, \qquad 
	1 \leq i \leq  N_\ell, \,\, 
	0 \leq \ell \leq L, \,\, 0 \leq \scaleparam \leq S.
\end{equation}
Here, $t_{\ell,\scaleparam} $ is a reference time-to solution per
sample on level $\ell$. Several natural choices exist, such
as the mode, median, mean or minimum over a sample set. The term 
$C_{\ell,\scaleparam}(\omega^{i})$ encapsulates fluctuations
across samples. It depends on the robustness of the PDE solver, as
well as on the type of parallel computer system.
It is scaled such that it is equal to one if there are no run-time variations.
Fig. \ref{fig:strongam} (right)
shows a typical run-time distribution for $2048$ samples 
each of which was computed on $512$ processors
with $\ell=0$ and $\sigma^2 =0.5$ in \eqref{def:expcov}. 

Assuming no efficiency loss due to load imbalances and an optimal parallel 
efficiency for $\scaleparam=0$, the theoretical optimal mean run-time 
for the MLMC method~is
\begin{equation}
t\foot{mlmc}^{\text{opt}} =  \frac{P_0^{\min}}{P\foot{max}} \, \sum_{\ell=0}^{L} N_\ell 2^{3\ell}
\mathbb{E}(C_{\ell,0}) t_{\ell,0} \,  .
\label{eq:optruntime}
\end{equation}

There are three main sources of inefficiency in parallel MLMC
algorithms: (i) a partly idle
machine due to large run-time variations between samples
scheduled in parallel, 
(ii) non-optimal strong scalability properties of the solver, i.e.,
$t_{\ell,\scaleparam} > 2 t_{\ell,\scaleparam-1}$,  or (iii) over-sampling, i.e., more samples than 
required are scheduled to fill the machine. In the following we address (ii) and (iii) in more detail.

The strong parallel efficiency of a solver can be charaterized in terms of 
$\text{Eff}_\ell(\scaleparam) := t_{\ell,0} /(2^\scaleparam  t_{\ell,\scaleparam})$.
In order to predict $t_{\ell,\scaleparam}$, $1 \leq \scaleparam \leq S$,
we define a surrogate cost function depending on $0 \leq \scaleparam \leq S
$ that is 
motivated by Amdahl's law \cite{hager2010introduction}:
\begin{equation}
\label{eq:amdahls}
t_{\ell,\scaleparam} \approx t_{\ell,0} (B + 2^{-\scaleparam} (1-B)) , \quad 
\text{Eff}_\ell (\scaleparam) \approx (2^\scaleparam B + (1-B))^{-1}
 .
\end{equation}
The serial fraction parameter $B$ in \eqref{eq:amdahls} quantifies the
amount of non-parallelizable work. It 
can be 
calibrated from time measurements.
For a solver with good scalability properties, $B$ 
is almost constant over the levels
so that we use a single value on all levels. 
Fig.\ \ref{fig:strongam} (left) shows the typical range of 
the scalability window, i.e., $S=4$, for $\ell =0$ and $P_0^{\min} =512$.
We also see the influence of different serial fraction parameters
$B \in \{ 0,0.01, 0.1,1\}$ on the parallel efficiency
and the good agreement of the cost model (\ref{eq:amdahls}) with 
averaged measured run-times. The fitted serial
fraction parameter $B$ lies in the range of $[0.01, 0.03]$
for different types of PDE within the HHG framework. In an adaptive 
strategy, we can also use performance measurements 
from past computations to fit better values of $B$ in the cost
predictions for future scheduling steps. 

Let $J_\ell(\theta) \in \mathbb{N}$ denote the number of samples 
that can be at most computed simultaneously on level $\ell$ if 
$ 2^{3\ell+ \theta} P_0^{\min}$ processors are used 
per sample, and by $k_\ell^{\text{seq}} (\scaleparam)$
we denote the number of required sequential cycles to run in total a
minimum of $N_\ell$ samples.
 Then, $ J_\ell(\theta)$,  $k_\ell^{\text{seq}} (\scaleparam)$ and the associated relative load imbalance $\text{Imb}_\ell (\scaleparam)$ are given by
\begin{equation} \label{eq:seqsteps} 
J_\ell(\theta) =  \left\lfloor\frac{P\foot{max}}{2^{3\ell+ \theta} P_0^{\min}}\right\rfloor, \quad 
k_\ell^{\text{seq}} (\scaleparam) = \left\lceil
  \frac{N_\ell}{J_\ell(\scaleparam)} \right\rceil,
\quad \text{Imb}_\ell (\scaleparam) := 1-  \frac{2^{3\ell
    +\scaleparam} P_0^{\min} N_\ell}{
 k_\ell^{\text{seq}} (\scaleparam) P\foot{max}} .  \end{equation}
We note  that $ 0 \leq \text{Imb}_\ell (\scaleparam) < 1$, with $\text{Imb}_\ell (\scaleparam) =0$ when no load imbalance occurs. For $\text{Imb}_\ell (\scaleparam)
> 0$,  part of the machine will be idle either due to the
$P\foot{max}/(2^{3\ell+ \theta} P_0^{\min}) \not\in \mathbb{N}$ or due
to $ N_\ell / J_\ell (\scaleparam) \not\in \mathbb{N}$. 

 The remaining processors in the last sequential steps can  be used to
compute additional samples that improve the accuracy,
but are not necessary to achieve the required tolerance, or we can
schedule samples on other levels in parallel (see the next section).  
The product 
\begin{equation} \label{eq:effs}
\eta_\ell (\scaleparam) := (1 - \text{Imb}_\ell(\scaleparam)) \text{Eff}_\ell
(\scaleparam) 
\end{equation} will be termed {\em MLMC level efficiency} and we note
that it also depends on $N_\ell$.

\subsection{Classification of concurrent execution strategies}
 We classify execution strategies
for MLMC methods in two ways, 
either referring to the layers of parallelisms or 
to the resulting time-processor diagram. 
\subsubsection{Layers of parallel execution}
\label{subsec:parallel_layers}
Especially on the finer grid levels in MLMC, the number of samples is too
small to fully exploit modern parallel systems by executing
individual samples in parallel. Multiple layers of parallelism 
must be identified. In the context of MLMC methods, 
three natural layers exist: 
 \begin{description}
\item[{\rm\em Level parallelism:}] The estimators
on level $\ell = 0,\ldots, L$ may be computed in parallel.
\item[{\rm\em Sample parallelism:}]
The samples $\{Y^i_{\ell}\}_{i=1}^{N_\ell}$ on level
$\ell$ may be evaluated in parallel.
\item[{\rm\em Solver parallelism:}]
The PDE solver to compute sample $Y_\ell^i$ may be parallelized.
\end{description}

The loops over the levels and over the samples are inherently parallel,
except for some minimal postprocessing to compute the statistical
quantities of interest. The challenge is how to balance
the load between different levels of parallelism and how to schedule 
the solvers for each sample. Especially in the adaptive setting, 
without a priori information, an exclusive use
of level parallelism is not always possible, but in most
practical cases, a minimal number of required levels 
and samples is known a priori. For the moment, we assume $L$ and 
$N_\ell $, $ 0 \leq \ell \leq L$, to be fixed and given.
In general, these quantities have to be determined
dynamically (cf.~Alg.~\ref{adaptive_mlmc_algorithm}).

The concurrent execution can now be classified according to
the number of layers of parallelism that are exploited: one, two, or three.
Typically, $P\foot{max} >
\sum_{l=0}^L N_\ell$ and $P\foot{max} \gg N_L$ on modern
supercomputers, and thus solver parallelism is mandatory 
for large-scale computing. Thus, the only possible one-layer approach
on supercomputers is the solver-only strategy. For a two-layer
approach, one can either exploit the solver and level layers or
the solver and sample layers. Since the number of levels~$L$ is,
in general, quite small, the solver-level strategy has significantly lower 
parallelization potential than a solver-sample strategy. 
Finally, the three-layer approach 
takes into account all three possible layers of parallelism and is 
the most flexible one.

\subsubsection{Concurrency in the processor-time diagram}

\begin{figure}[ht]
\centering
\def\svgwidth{0.6\textwidth}
{\relsize{-2} 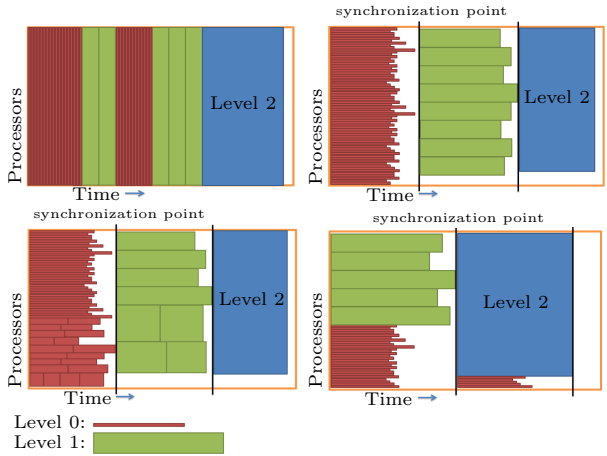}
\caption{\label{fig:seqhomscheduling} Upper row: illustration of 
 homogeneous bulk synchronous strategies; one-layer
(left) and two-layer parallelism (right); Lower row:
illustration of heterogeneous bulk synchronous strategies.
 two-layer 
(left) and three-layer parallelism (right).}
\end{figure}

An alternative way to classify different parallel execution models is to
consider the time-processor diagram, where the scheduling of each 
sample $Y^{i}_{\ell}$, $1 \leq i \leq N_\ell$, $ 0 \leq \ell \leq L$,
is represented by a rectangular box with the height representing the 
number of processors used. A parallel execution model is called
{\em homogeneous bulk synchronous} if 
at any time in the processor diagram,
all tasks execute on the same level with the same number of processors.
Otherwise it is called
{\em heterogeneous bulk synchronous}. 
The upper row of Fig.~\ref{fig:seqhomscheduling} illustrates two examples of
homogeneous bulk synchronous strategies, whereas
the lower row presents two heterogeneous strategies.

The one-layer homogeneous strategy, as shown in Fig. 
\ref{fig:seqhomscheduling} (left), offers no 
flexibility. The theoretical run-time is simply given by $\sum_{\ell=0}^L
\sum_{i=1}^{N_\ell} t(i,\ell, \theta_l^{\max})$, where
$\theta_\ell^{\max}$ is such that $P\foot{max} = 
2^{3\ell+\theta_\ell^{\max}} P_0^{\min} $.  It guarantees perfect load balancing,
but will not lead to a good overall efficiency 
since on the coarser levels $\theta_\ell^{\max} $ is typically significantly larger than $S$.  On the coarsest level we may even have $M_0 <
P\foot{max}$, i.e., less grid points than processors. Thus we will not further 
consider this option.

\section{Examples for scheduling strategies} \label{sec:impl}
Our focus is on scheduling algorithms that
are flexible with respect to the scalability window of the PDE solver
and robust up to a huge number of processors $P\foot{max}$.
To solve the optimization problems, we will either impose additional 
assumptions that allow an exact solution,
or we will use meta-heuristic search algorithms such as, e.g.,
simulated annealing  \cite{van1987simulated,van1992job}.
Before we introduce our scheduling approaches, we comment briefly 
on technical and practical aspects that are important for the implementation.

\paragraph{Sub-communicators}
To parallelize over samples as well as within
samples, we split the MPI\_COMM\_WORLD communicator via the
MPI\_Comm\_split command and provide each sample with its 
own MPI sub-communicator.
This requires only minimal changes to the multigrid algorithm and
all MPI communication routines can still be used. 
A similar approach, using the MPI group
concept, is used in~\cite{vsukys2012static}.

\paragraph{Random number generator}
To generate the samples of the diffusion coefficient $k(x,\omega)$ we
use the approach described in Sect.\ \ref{sec:random-fields}.
This requires suitable random numbers for the definition of the white
noise on the right hand side of \eqref{spde}. For large scale MLMC computations
we select the \textit{Ran}~\cite{press2007numerical} generator
that has a period of $\approx 3.1 \cdot 10^{57}$ and is thus suitable even
for $10^{12}$ realizations.
It is parallelized straightforwardly
by choosing different seeds for each process,
see, e.g.,~\cite{leva1992fast}.   

We consider now examples for the different classes of 
scheduling strategies. 

\subsection{Sample synchronous and level synchronous homogeneous}
\label{sec:SaSyHom}
Here, to schedule the samples we assume that the run-time of the solver 
depends
on the level $\ell$ and on the number of associated processors, but
not on the particular sample $Y^i_{\ell}$, $ 1 \leq i \leq N_\ell$.
As the different levels are treated sequentially and each
concurrent sample is executed with the same number of processors, we
can test all possible configurations. 

Let $0 \le \ell \le L$ be
fixed. Then, for a fixed $0 \leq \scaleparam \leq S$, the total time
on level $\ell$ is $k_\ell^{\text{seq}} (\scaleparam)
t_{\ell,\scaleparam}\,$. We select the largest index 
$\scaleparam_\ell \in \{0,1, \ldots, S\}$ such that
$$
\scaleparam_\ell
= \arg \min_{0 \leq \scaleparam \leq S} 
k_\ell^{\text{seq}} ( \scaleparam) t_{\ell,\scaleparam} =
\arg \max_{0 \leq \scaleparam \leq S} 
\text{Eff}_\ell ( \scaleparam) (1- \text{Imb}_{\ell} (\scaleparam))
.$$
Thus the minimization of the run-time per level is equivalent to 
a maximization of the total level efficiency.
The computation of $\scaleparam_\ell$ is trivial provided $
t_{\ell,\scaleparam}$ is known for all $\scaleparam$. We can either
set it to be the average of pre-computed timings of the solver on
level $\ell$ or we can use  \eqref{eq:amdahls} with a fitted serial 
fraction parameter $B$. In that case,
$$
\scaleparam_\ell = \arg\min_{0 \leq \scaleparam \leq S} k_\ell^{\text{seq}} ( \scaleparam) (B + 2^{-\scaleparam} (1-B)) .
$$
The level $\ell$ only enters this formula 
implicitly, through $N_\ell$ and through the growth factor $2^{3\ell}$.
Given $\scaleparam_\ell$, we can group the processors accordingly and run
$k_\ell^{\text{seq}} ( \scaleparam_\ell) $ sequential steps for each
level $\ell$. Note that the actual value of
$t_{\ell,0} $ does not influence the selection of $\scaleparam_\ell $\,. It does of course influence the
absolute run-time.

\begin{figure}[t]
 \includegraphics[trim= 0mm 40mm 0mm 10mm,width=0.325\textwidth]{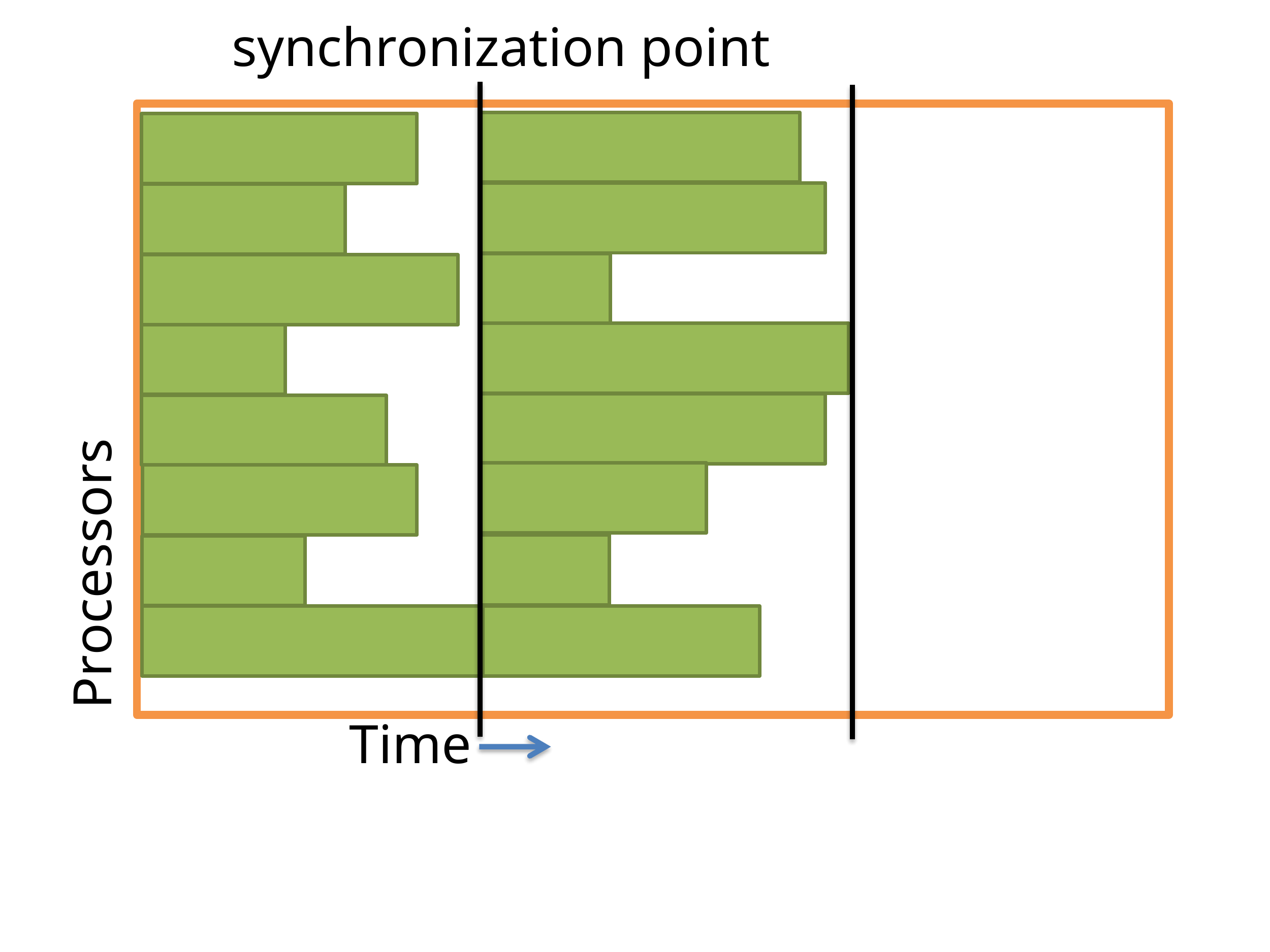}
\includegraphics[trim= 0mm 40mm 0mm 10mm,width=0.325\textwidth]{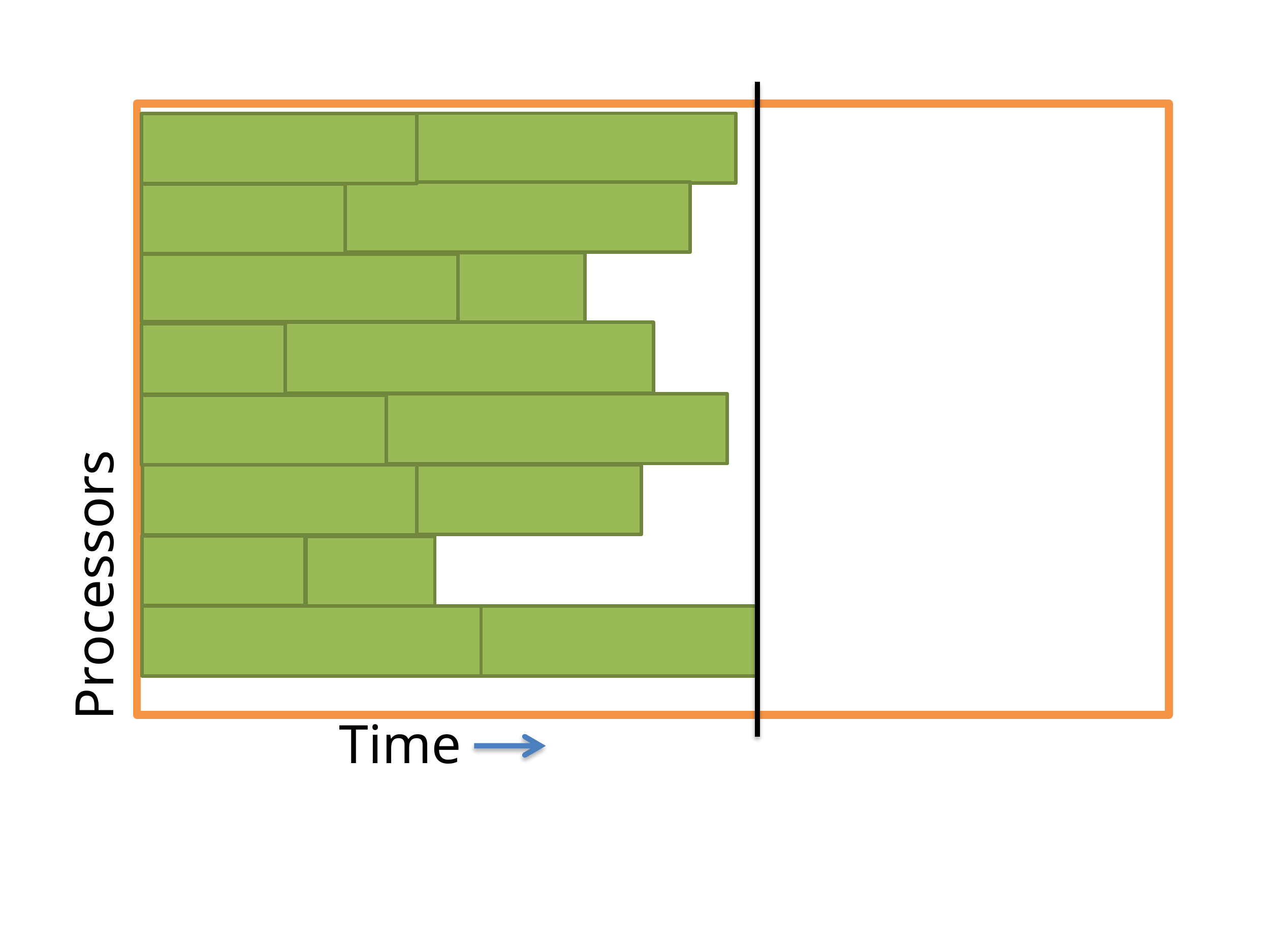}
  \includegraphics[trim= 0mm 40mm 0mm 10mm,width=0.325\textwidth]{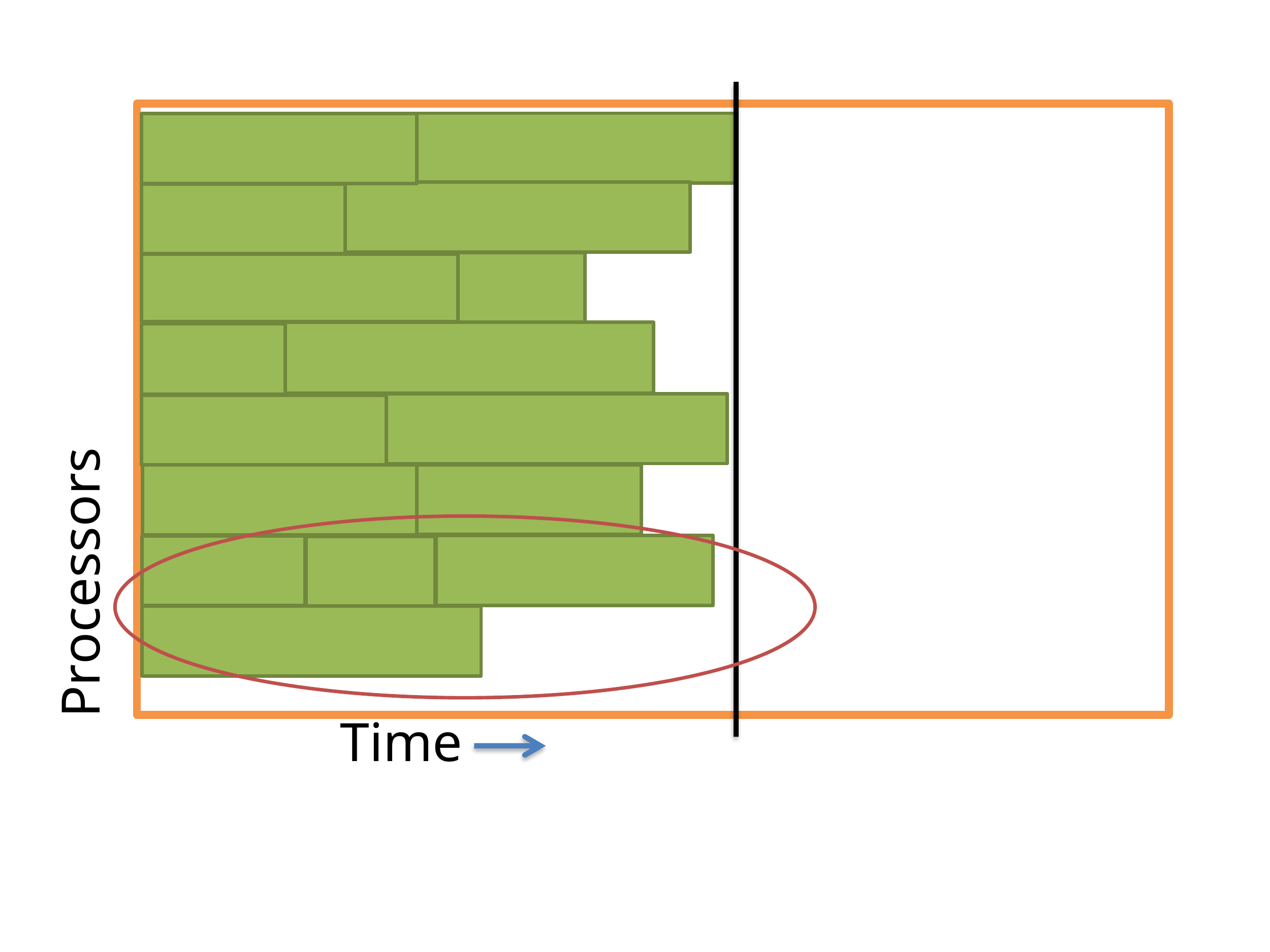}
\caption{\label{fig:diff} Illustration of different homogeneous
  scheduling strategies. Left: sample synchronous homogeneous
  (SaSyHom); Centre: level synchronous homogeneous (LeSyHom); Right:
  dynamic level synchronous homogeneous (DyLeSyHom, Sec.~\ref{sec:dynam}).}
\end{figure}

We consider two variants: (i) {\em Sample synchronous homogeneous (SaSyHom)}
imposes a synchronization step after each sequential step (see Fig.\
\ref{fig:diff}, left). Here statistical quantities can be updated after
each step. (ii) {\em Level synchronous homogeneous (LeSyHom)}, where
each block of $2^{3\ell+ \theta} P_0^{\min}$ processors executes all $k_\ell^{\text{seq}}
(\scaleparam_\ell) $ without any synchronization (see Fig.\
\ref{fig:diff}, centre).
Altogether 
$k_\ell^{\text{seq}} (\scaleparam_\ell)  J_\ell( \scaleparam_\ell)
\geq N_\ell$ samples are computed. When the run-time does not vary
across samples, both strategies will results in the same MLMC
run-time. If it does vary then the LeSyHom strategy
has the advantage that 
fluctuations in the run-time $t(i, \ell, \scaleparam)$ 
will be averaged and a shorter overall MLMC run-time can be expected
for sufficiently large
$k_\ell^{\text{seq}} (\scaleparam_\ell)$.

\subsection{Run-time robust homogeneous} \label{sub:rtr}
So far, we have
assumed that the run-time is sample independent, which is
idealistic (see Fig.~\ref{fig:strongam}, right). In the experiment in
Fig.~\ref{fig:strongam} (right), 3 out of 2048 samples required a
run-time of $50s$ on $P_0^{\min}=512$ processors. On a 
large machine with $P\foot{max} =524 288$ and with $\theta_0 =0$ we
need only $k_0^{\text{seq}}(0)=2$ sequential steps on level $0$. Therefore, the
(empirical) probability that the SaSyHom strategy leads to a runtime of $100s$ is 
about $75 \%$, while the theoretical optimal run-time is
$\frac{2}{2048} \sum_{i=1}^{2048} t_i \approx 90s$. Here, $t_i $ is
the actual run-time of the ith sample from Fig.~\ref{fig:strongam} (right).
The probability that the LeSyHom strategy leads to a runtime of
$100s$ is less than $1\%$; in all other cases,  a
run-time of  $\leq 96s$ is achieved.

Let us now fix $0 \le \ell \le L$ again and include run-time
variations in the determination of $\scaleparam_\ell$.
Unfortunately, in general, run-time distribution functions are not
known analytically,
and thus the expected run-time
\begin{equation} \label{eq:maxr}
 E_{\ell,\scaleparam}  := \mathbb{E} \left[\max_{1 \leq j \leq J_\ell (\scaleparam)} \Bigg(
\sum_{k=1}^{k_\ell^{\text{seq}} (\scaleparam) } t( i_{jk}, \ell, 
\scaleparam)  \Bigg) \right]
\end{equation}
cannot be computed explicitly. Here, the samples
are denoted by $i_{jk}$ with $j = 1,\ldots, J_\ell (\scaleparam)$ and 
$k =1, \ldots, k_\ell^{\text{seq}} (\scaleparam)$, related to their
position in the time-processor diagram.
The expression in \eqref{eq:maxr} yields the actual, expected run-time on level
$\ell$ to compute $J_\ell (\scaleparam)
k_\ell^{\text{seq}} (\scaleparam) \geq N_\ell$ samples with 
$ 2^{3\ell+\scaleparam} P_0^{\min}$ processors per sample
when no synchronization after the sequential steps is
performed.

The main idea is now to compute an approximation
$\widehat{E}_{\ell,\scaleparam}$ for 
$E_{\ell,\scaleparam}$, and
then to minimise $\widehat{E}_{\ell,\scaleparam}$ for each $\ell$,
i.e., to find  $\scaleparam_\ell$ such that
$\widehat{E}_{\ell,\scaleparam_\ell} \leq \widehat{E}_{\ell,\scaleparam}$ for all
$0 \leq \scaleparam \leq S$. As a first approximation, we replace $t(i_{jk}, \ell, 
\scaleparam)$ by the approximation $C_{\ell,\scaleparam}(\omega^{i_{jk}})
\, t_{\ell, \scaleparam}$ in~\eqref{eq:time-for-sample} and assume
that the  stochastic cost factor distribution
neither depends on the level $\ell$ nor on the scale parameter $\scaleparam$.
Furthermore, we approximate the expected value by an average
over $\grossemm$ samples to obtain the approximation
\begin{equation} \label{eq:maxrsM}
  \widehat{E}_{\ell,\scaleparam}(\grossemm) := \frac{1}{\grossemm}
  \sum_{m=1}^\grossemm \max_{1 \leq j \leq J_\ell(\scaleparam)} \left(
    \sum_{k=1}^{k_\ell^{\text{seq}} ( \scaleparam) } C_{0,0}(
    \omega^{i_{jkm}}) \right) t_{\ell,\scaleparam} \,,
\end{equation}
where $i_{jkm} := i_{jk} + (m-1) J_\ell(\scaleparam) k_\ell^{\text{seq}}(\scaleparam)$. If reliable
data for $t_{\ell,\scaleparam} $ is available we define
$$
\scaleparam_\ell := \arg\min_{0 \leq \scaleparam \leq S} \widehat{E}_{\ell,\scaleparam}(\grossemm)  
$$
Otherwise, we include a further approximation and
replace $t_{\ell,\scaleparam}$ by $B+2^{-\scaleparam} (1-B)$ in 
\eqref{eq:maxrsM} before finding the minimum of
$\widehat{E}_{\ell,\scaleparam}(\grossemm)$. Here, we still require 
an estimate for the serial fraction parameter $B$. To decide on the
number of samples $\mu$ in \eqref{eq:maxrsM}, we keep increasing $\mu$
until it is large enough so that $\grossemm$ and 
$\grossemm/2$ yield the same $\scaleparam_\ell$.  For all our test 
settings, we found that $\grossemm \leq 500$ is sufficient. 

To evaluate 
\eqref{eq:maxrsM},
we need some information on the stochastic cost factor
$C_{0,0}(\omega)$ which was assumed to be constant across levels and
across the scaling window.
We use a run-time histogram associated with level $\ell=0$. This
information is either available from past computations or can be built up
adaptively within a MLMC method. Having the run-times $t_k$, $1 \leq k 
\leq K$, of $K$ samples on level $\ell=0$ at hand, we emulate 
$C_{0,0}( \omega )$ by using a pseudo random integer generator from 
a uniform discrete distribution ranging from one to $K$ and replace
the obtained value $j \leq K$ by $t_j K /\sum_{k=1}^K t_k$.
%
Having computed the value of $\scaleparam_\ell$, we proceed as 
for LeSyHom and call this strategy {\em run-time robust homogeneous (RuRoHom)}.
For constant run-times, RuRoHom yields again the same run-times as
LeSyHom and as SaSyHom. 

\subsection{Dynamic variants}
\label{sec:dynam}
So far, we have used pre-computed values for $\scaleparam_\ell$
and $k_\ell^{\text{seq}} ( \scaleparam_\ell)$ in all variants, and each processor block carries
out the computation for exactly  $k_\ell^{\text{seq}} ( \scaleparam_\ell)$ samples.
For large run-time variations, this will still lead to unnecessary inefficiencies. 
Instead of assigning samples to each processor block a-priori,
they can also be assigned to the processor blocks dynamically
at run-time.
As soon as a block terminates a computation, 
a new sample is assigned to it until the required number $N_\ell$ is reached.
This reduces over-sampling and can additionally
reduce the total run-time on level $\ell$. However, on massively parallel
architectures this will only be efficient when the
dynamic distribution of samples
does not lead to a significant communication overhead.
The dynamic strategy can be combined with either the LeSyHom 
or the RuRoHom approach and we denote them 
{\em dynamic level synchronous homogeneous
(DyLeSyHom)} and {\em dynamic run-time robust homogeneous
(DyRuRoHom)}, respectively.
Fig.\ \ref{fig:diff} (right) illustrates the DLeSyHom strategy. 
Note specifically that here not all processor blocks 
execute the same number of sequential steps.

In order to utilize the full machine, it is crucial that no processor
is blocked by actively waiting to coordinate the asynchronous execution.
The necessary functionality may not be fully supported on all parallel systems.
Here we use the MPI~2.0 standard that permits
one-sided communication and thus allows
a non-intrusive implementation.  
The one-sided communication is accomplished by remote direct memory access 
(RDMA) using 
registered memory windows. 
In our implementation, we create a window on one processor to synchronize
the number of samples that are already computed.
Exclusive locks are performed on a get/accumulate combination to
access the number of samples.

%
%
\subsection{Heterogeneous bulk synchronous scheduling} 
\label{sec:heter}
Heterogeneous strategies are clearly more flexible than homogeneous
ones, but the number of scheduling possibilities grows exponentially.
Thus, we must first reduce the complexity of the scheduling problem.
In particular, we ignore again run-time variations and assume 
$t(i,\ell, \scaleparam) = t_{\ell,\scaleparam}$. 
We also assume that $N_\ell > 0$ on all levels $\ell = 0, \ldots, L$.
Within an adaptive strategy, samples may only be required on some of
the levels at certain times and thus this condition has to  hold true
only on a subset of ${\mathcal I} := \{0, \ldots , L \}$.

In contrast to the homogeneous setups, we do not aim to find scaling
parameters~$\scaleparam_\ell$ that minimize the
run-time on each level separately, but instead minimize the total MLMC
run-time. We formulate the minimization process as two
constrained minimization steps that are coupled only in one direction, where we
have to identify  the number  $N_{\ell,\scaleparam} \in \mathbb{N}_0$ of samples on level
$\ell$ which are carried out in parallel with $ 2^{3\ell +\scaleparam}
P_0^{\min}$ processors, as well as the number $
k_{\ell}^{\text{seq}}(\scaleparam)$
 of associated sequential steps. Firstly, assuming $N_{\ell,\scaleparam}$ to be given, for all $ 0 \leq \ell \leq L$, $0 \leq
\scaleparam \leq S$, we solve the constrained minimization problem for $k_{\ell}^{\text{seq}}(\scaleparam)$
$$
\arg \min_{k_{\ell}^{\text{seq}}(\scaleparam)  \in \mathbb{N}_0} \left( \max_{0 \leq \scaleparam \leq S} 
  t_{\ell,\scaleparam} \, k_{\ell}^{\text{seq}}(\scaleparam) \right) , \qquad
\sum_{\scaleparam=0}^S N_{\ell,\scaleparam} \, k_{\ell}^{\text{seq}}(\scaleparam) \geq N_\ell .
$$
Secondly, having $k_{\ell}^{\text{seq}}(\scaleparam)$ at hand, we find
values for $N_{\ell,\scaleparam} \in \mathbb{N}_0$ such as to 
minimize
$$
\arg \min_{N_{\ell,\scaleparam} \in \mathbb{N}_0} 
\max_{ 0 \leq \scaleparam \leq S \atop \, 0 \leq \ell \leq L}
t_{\ell,\scaleparam} k_{\ell}^{\text{seq}}(\scaleparam),
$$
 the expected run-time, subject to the following inequality  constraints  
\begin{subequations}
\label{eq:constraints}
\begin{align}
 0 \leq N_{\ell,\scaleparam} \leq 2^{-3\ell} 2^{-\scaleparam} P\foot{max} /P_0^{\min}
\label{eq:constraint0}, \\
 \sum_{\scaleparam=0}^S  N_{\ell,\scaleparam} > 0 , \text{ for } \ell \in \mathcal{I}, \label{eq:constraint1}\\
\sum_{\ell=0}^L  \sum_{\scaleparam=0}^S  N_{\ell,\scaleparam}2^{3\ell} 2^\scaleparam P_0^{\min} \leq P\foot{max}
\label{eq:constraint2}.
\end{align}
\end{subequations}

We apply integer 
encoding~\cite{srinivas1994genetic} for the 
initialization and for possible mutation operators to guarantee that 
$N_{\ell,\scaleparam} \in \mathbb{N}_0$. Clearly, if $N_{\ell,\scaleparam} \in \mathbb{N}_0$ then \eqref{eq:constraint2}
implies \eqref{eq:constraint0}. However, even though it is redundant, 
\eqref{eq:constraint0} is 
enforced explicitly to restrict 
the search space in the meta-heuristic optimization algorithm. 
The condition in \eqref{eq:constraint1} that at least
one sample is scheduled on each level at
all times could also be relaxed. However, this would require a 
redistribution of processors in the optimization problem and can 
significantly increase the algorithmic and technical complexity. 
If \eqref{eq:constraint1} is violated on some level $\ell$,
we set $N_{\ell,0} = 1$. Condition \eqref{eq:constraint2}, however, is
a hard constraint. The
number of processors that are scheduled cannot be larger than
$P\foot{max}$. If \eqref{eq:constraint2} is violated, we enforce it by 
a repeated multiplication of $N_{\ell,S} , \ldots , N_{\ell,0}$ by $1/2$ until it holds. 
 At first glance this possibly leads to an 
 unbalanced work load,
 but the applied  meta-heuristic search strategy compensates for it.
%
With the values of $N_{\ell,\scaleparam}$ identified, the samples are 
distributed  dynamically onto the machine, see also \cite{vsukys2014adaptive}.

To illustrate the complexity of this optimization task, we consider the number 
of different combinations for $N_{\ell,\scaleparam}$ 
that satisfy \eqref{eq:constraint0} but not necessarily \eqref{eq:constraint1} and \eqref{eq:constraint2}.
For example, for $L=3$, $S=4$, $P\foot{max}=8\,192$ and
$P_0^{\min}=1$, there are $\mathcal{O} (10^{39})$ possible combinations.
Even for the special case that the scalability window degenerates,
i.e., that $S=0$, there are still $\mathcal{O} (10^{10}) $ possibilities.

As an example for the following two subsections, we consider $(N_0,N_1,N_2,N_3) = (4123, 688, 108, 16)$
and  actual run-times from measurements in a set of numerical
experiments:
{\small \begin{equation} \label{eq:matrixss}
( t_{\ell,\scaleparam})_{0 \leq \ell \leq 3 , \atop 0 \leq \scaleparam \leq 4} = 
\begin{pmatrix}
167 & 83.84 & 42.30 & 21.63 & 11.60 \\
 171 & 86.28 & 44.53 & 23.13 & 12.41 \\
 177 & 90.40 & 47.07 & 24.21 & 12.97 \\
 179 & 91.61 & 48.27 & 24.86 & 13.63
\end{pmatrix} .
\end{equation}}

\subsubsection{The degenerate case $S=0$ and a new auxiliary objective}
For $S=0$, a cheap but non-optimal way to choose $N_{\ell,0} $ is
\begin{equation}
N_{\ell,0}  =  \left\lfloor \frac{P\foot{max} \, N_{\ell} \, t_{\ell,0} }{ \sum_{i=0}^{L}  N_{i} 2^{3i} P_0^{\min}   t_{i,0} } \right\rfloor \,.
\label{eq:simpleestimation}
\end{equation}
The corresponding run-time is 
$\max_{\ell=0, \ldots , L}  t_{\ell,0} \lceil N_\ell /N_{\ell,0}
\rceil$. The total number of 
processors is $\sum_{\ell=0}^L N_{\ell,0} 2^{3\ell}
P_0^{\min}$. This choice is acceptable when the workload is
evenly distributed across levels, which is one of the typical scenarios in MLMC.
It also requires that the full machine can be exploited
without any imbalance in the workload.

\begin{figure}[ht]
\centering
\resizebox{0.28\textwidth}{!}{\relsize{5} 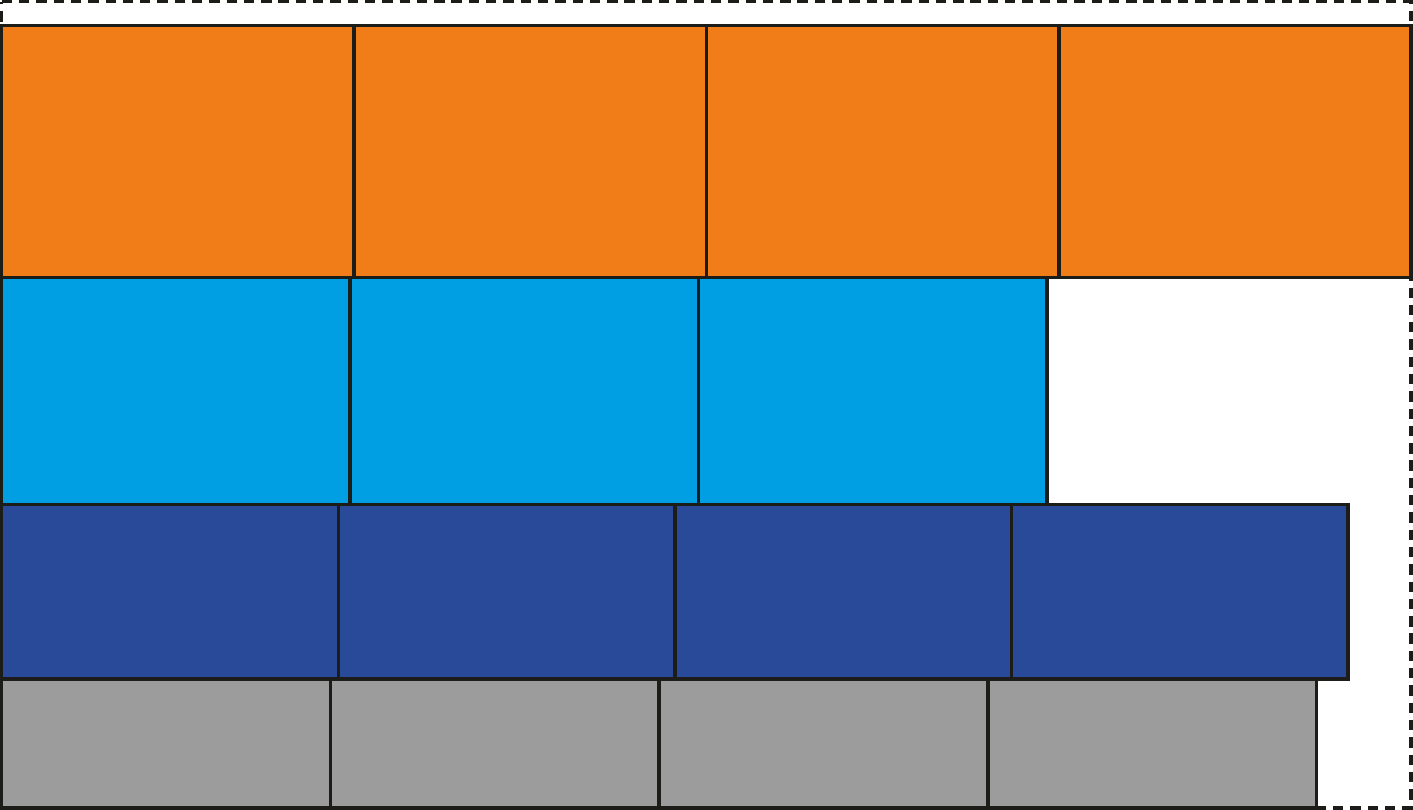}
\hspace*{1cm}
\resizebox{0.28\textwidth}{!}{\relsize{5} 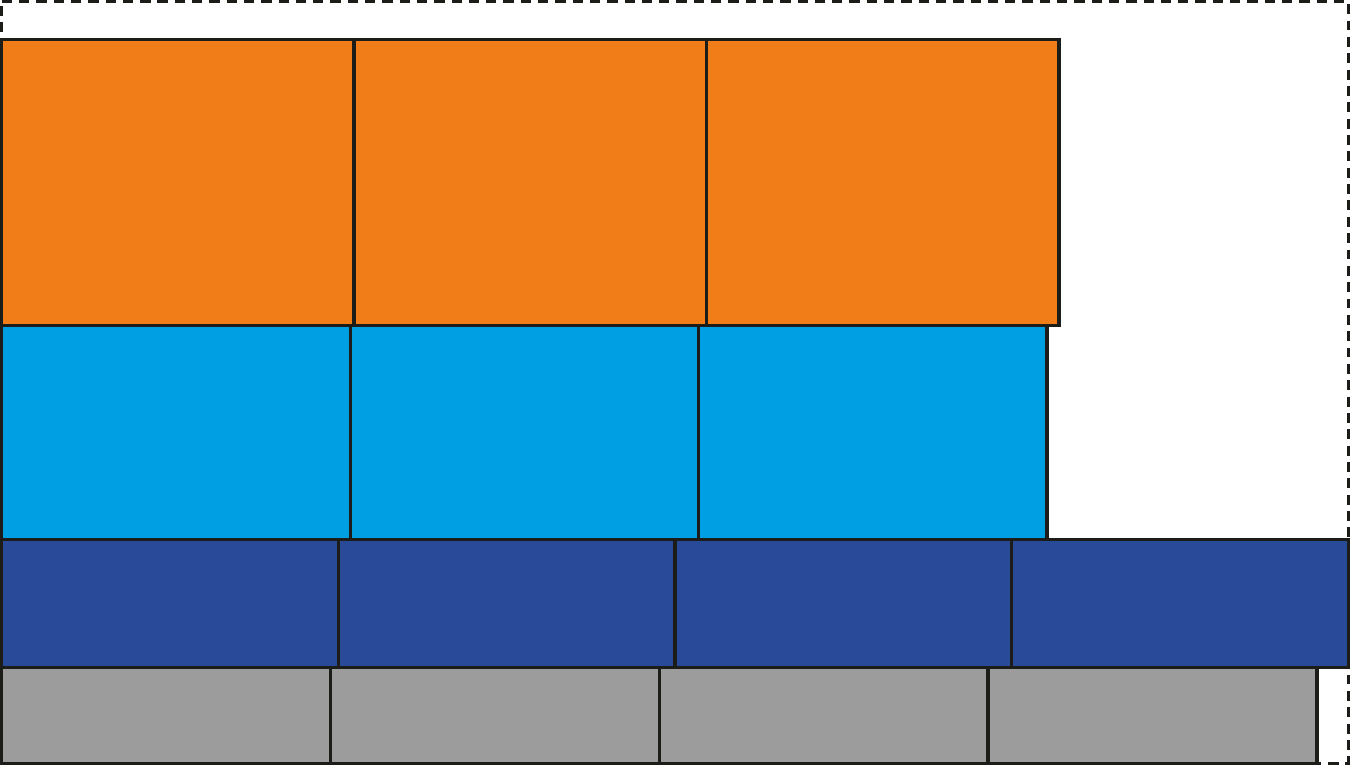}
\caption{\label{fig:simplesched} Different scheduling patterns: Selection of $N_{\ell,0} $ by \eqref{eq:simpleestimation} (left) and optimal choice of $N_{\ell,0} $}
\end{figure}

Using  the first column of \eqref{eq:matrixss} in \eqref{eq:simpleestimation}, we find  as total run-time $\SI{716}{\second}$ and the distribution
  $(N_{0,0}, N_{1,0}, N_{2,0}, N_{3,0}) = (1314, 221, 36, 5)$, see the
  left of Fig.~\ref{fig:simplesched}, while on the right, the minimal
  run-time pattern with $\SI{684}{\second}$ is illustrated.
Due to weak scaling effects, the lower levels tend to have a larger number of
sequential steps than the higher ones.
We note that there exist many different configurations such that the minimal run-time is reached. Using \eqref{eq:simpleestimation} as starting guess, and then performing a local adaptive neighborhood search  is much cheaper than an exhaustive search.

However for $S> 0$, we cannot define a good starting guess as easily
and have to resort to meta-heuristic strategies. We consider
simulated annealing (SA) techniques, see, e.g., \cite{van1987simulated,Weg09}, 
which provide a computationally feasible approach to solve complex
scheduling problems approximately. We start with $S=0$.
The following experiments were performed with Python 
using \textit{inspyred}\footnote{Garrett, A. (2012). inspyred (Version 1.0). Inspired Intelligence Initiative. Retrieved from http://github.com} with minor modifications. 
The temperature parameter in the  SA method is decreased using a
geometric schedule $T_{k+1} = 0.8 \, T_k$. The initial temperature is
chosen to be $T_0 = 10^{3}$ which is of the order of the initial
changes of the objective function. 

\begin{figure}[ht]
\centering
  \includegraphics[trim= 0mm 0mm 0mm 0mm,width=0.8\textwidth]{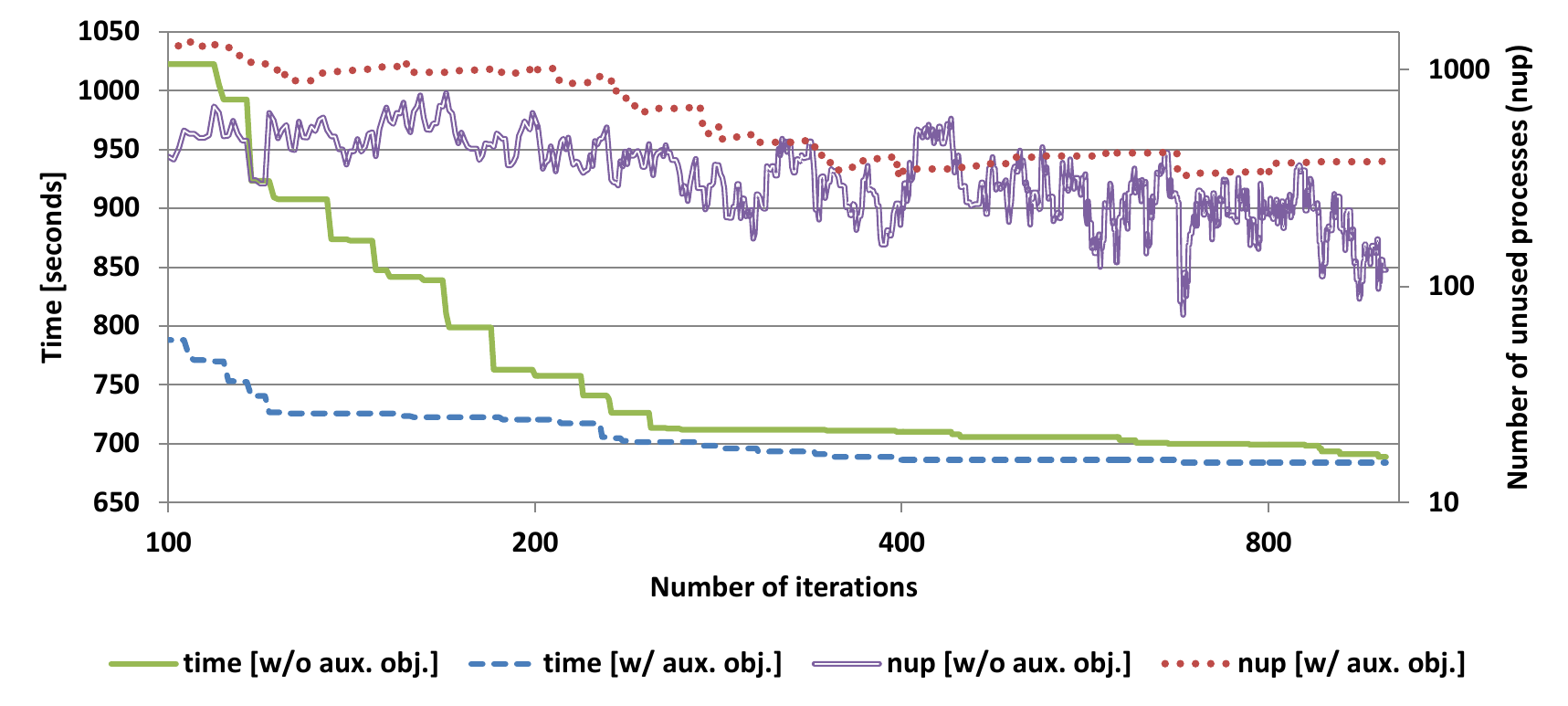}
\caption{\label{fig:auxobj}Average MLMC run-time and number of unused processors (nup) with and without auxiliary objective (w/ and w/o aux. obj.) w.r.t. the number of SA iterations.}
\end{figure}

Here we  choose a Gaussian mutation with distribution $\mathcal{N}(0, 0.1\, P\foot{max} / ( 2^{3\ell} P_0^{\min}))$ and a mutation rate of $0.2$ guaranteeing that roughly  one gene per SA iteration is changed. All runs  were repeated ten times with different seeds and we report
minimal (min), maximal (max), as well as the arithmetically averaged
(avg) MLMC run-times. Selecting $ 1\,000$ evaluations as the stopping
criterion in SA, we obtain $t[\min,\text{avg},\max] = [684, 691.2, 708]\,\si{\second}$. 
For comparison, a stopping criterion of $2\,000$ evaluations yields 
$t[\min,\text{avg},\max] = [684, 684, 684]\,\si{\second}$.  
Fig.~\ref{fig:auxobj} 
shows the evolution of the average MLMC 
run-time between iteration 100 and 1\,000 in the SA. (Please refer to
the curve labelled ``time [w/o aux.\ obj.]''). Minimizing only the
run-time in the SA objective function, we observe that between
iteration  $250$ and iteration $800$ almost no decrease in the average run-time
is achieved.
This is due to the rather flat structure of the objective function
in large parts of the search domain
resulting from the fact that 
many different possible combinations yield  identical run-times. 

To improve the performance of the SA scheduling optimizer, 
we introduce the number of idle processors as 
a second auxiliary objective. 
This auxiliary objective is only considered, if two candidates result in
the same MLMC run-time.
In this case the candidate with the higher number of idle processors is selected.
This choice is motivated by the observation that the probability to find a candidate
with shorter run-time is higher in the neighborhood of a candidate that has 
more idle processors. Fig.~\ref{fig:auxobj} shows clearly
that the optimization can be accelerated by including the
auxiliary objective.
Here, we find a MLMC run-time of less than $\SI{700}{\second}$ in less than $300$
iterations.
The optimal run-time can be obtained with $(N_{0,0}, N_{1,0}, N_{2,0}, N_{3,0}) = (1031, 172, 36, 6)$ and a total of $7783$ processors used.
From now on, we always include the number of idle processors as a secondary objective
 in the SA optimization algorithm.

\subsubsection{The highly scalable case $S=4$ and new hybrid mutants}
We use the example data in \eqref{eq:matrixss} and compare five different mutation operators.
In addition to the already considered Gaussian mutation, we also use simpler
and more sophisticated strategies.
Random reset mutation replaces a gene by a uniform randomly
chosen integer satisfying  \eqref{eq:constraint0}. 
In the case of non-uniform mutation, see \cite{michalewicz1996genetic},
a variation is added to the selected gene and the mutation depends on the SA step. 
The  initial mutation strength is set to one and decreases with increasing
iteration numbers. 
Tab.\ \ref{tab:mutoperators} shows that the Gaussian mutation
is superior to both the random reset as well as the non-uniform mutation.
However, even for the Gaussian mutation, more than $50\, 000$ SA
iterations are necessary to find average run-times close to the optimal one.%
\begin{table}[h]
\caption{Comparison of obtained MLMC run-times
 (min, avg, max) for different mutation operators and SA  iteration numbers.}
\footnotesize
\begin{tabular}{lrrrr}
\hline
Mutation                & 1\,000               & 4\,000              & 16\,000             & 64\,000             \\ \hline
Random reset                 & $627.6,676.4,772.4$ & $624.6,641.0,676.8$   & $627.6,633.5,659.0$ & $603.9,616.4,632.8$                  \\
Non-uniform              & $641.2,717.0,774.6$  &$ 627.6,673.2,717.1$   & $627.6,647.4,716.0$   & $627.6,638.5,641.2$ \\
Gaussian                & $612.0,633.9,641.2$  & $605.2,624.8,635.5$ & $603.9,614.6,624.6$ & $603.9,608.0,612.0$ \\
Hybrid A      & $624.6,632.7,641.2$  & $603.9,608.0,612.0$ & $604.5,604.5,604.5$ & $604.5,604.5,604.5$ \\
Hybrid B  & $603.9,619.8,627.3$  & $603.9,603.9,603.9$ & $603.9,603.9,603.9$ & $603.9,603.9,603.9$ \\ \hline
\end{tabular}
\label{tab:mutoperators}
\end{table}

Thus, new problem-adapted mutation operators in the SA are essential.
We propose two new hybrid variants. 
Both perform first a Gaussian mutation and then a
problem adapted mutation, taking into account the required processor numbers.
The mutation rate for both is set to $0.1$. 

{\em Hybrid A:} 
In each step, we select randomly  two different ``genes''
$N_{\ell_1,\scaleparam_1}$ and $N_{\ell_2,\scaleparam_2}$, $0 \leq
\ell_1,\ell_2 \leq L$, $0 \leq \scaleparam_1,\scaleparam_2 \leq S$, as
well as a uniformly distributed random number $k \in [0,..,N_{\ell_1,\scaleparam_1} -1]$.
Then we mutate 
\begin{align*}
N_{\ell_1,\scaleparam_1} = N_{\ell_1,\scaleparam_1} -k \quad
\text{and} \quad
N_{\ell_2,\scaleparam_2} = N_{\ell_2,\scaleparam_2} + \left\lfloor k\, 2^{\scaleparam_1 -\scaleparam_2} 2^{3(\ell_1-\ell_2)} \right\rfloor .
\end{align*}
If  the original values for $N_{\ell_1,\scaleparam_1}$ and $N_{\ell_2,\scaleparam_2}$ were
admissible, satisfying the constraints \eqref{eq:constraints}, then
the mutated  genes are also admissible.
This type of mutation exploits the scalability window of the solver as
well as level parallelism.
For the special case $S=0$, it reduces to balancing the workload on
the different levels, by exploiting
the weak scalability of the solver. 

{\em Hybrid B.} This variant is proposed for a PDE solver 
that has a large scalability window. It follows the same steps, but
keeps $\ell_1 = \ell_2$ fixed, therefore only exploiting the strong and weak
scaling properties of the solver, but not the MLMC 
hierarchy. 

In Tab.\ \ref{tab:mutoperators}, we see that Hybrid B  shows the best
performance. Compared to Hybrid A it is less sensitive to the initial
guess and robustly finds a very efficient scheduling scheme  in less 
than $4 \, 000$ SA iterations. Thus,
we restrict ourselves to SA with Hybrid B type mutations in the 
following examples.
In the example considered in this section, it leads to the schedule
\begin{equation} \label{eq:matrixssr}
( N_{\ell,\scaleparam})_{0 \leq \ell \leq 3 , \atop 0 \leq \scaleparam \leq 4} = 
\begin{pmatrix}
0 &443  & 73 &0  &0  \\
 1 &98  & 0 &0  & 0 \\
0  &0  &3  & 3 &0  \\
 6 &0  &0  &0  &0 
\end{pmatrix} \quad 
( k_{\ell,\scaleparam}^{\text{seq}})_{0 \leq \ell \leq 3 , \atop 0 \leq \scaleparam \leq 4} = 
\begin{pmatrix}
 0& 7 &14  &0  &0  \\
 3 &7  &0  &0  &0  \\
 0 & 0 & 12 &24  &0  \\
 3 &0  &0  &0  &0 
\end{pmatrix}.
\end{equation}
Comparing the two cases $S=0$ and $S=4$ shows how important the strong scalability 
of the solver is to reach shorter MLMC run-times. It allows to reduce
the run-time by more than $10\%$, and thus the parallel MLMC
performance can be improved significantly with such an advanced
scheduling strategy. In the following, we call the scheduling
strategy {\em StScHet}, if strong scaling is included ($S >
0$). Otherwise, if no strong scaling is included ($S=0$), we call the scheduling strategy {\em
  noStScHet}.

To finish this section we summarise all the considered schedules
in Tab.\ \ref{tab:summary}.
\begin{table}[h]
\caption{\label{tab:summary}
Summary of parallel scheduling strategies.}
\centering
\begin{tabular}{cll}
\hline
Abbreviation & Schedule & Defined in \\ \hline
SaSyHom & Sample Synchronous Homogeneous & Sec.~\ref{sec:SaSyHom}\\
LeSyHom & Level Synchronous Homogeneous & Sec.~\ref{sec:SaSyHom}\\
RuRoHom & Run-Time Robust Homogeneous & Sec.~\ref{sub:rtr}\\
DyLeSyHom & Dynamic Level Synchronous Homogeneous & Sec.~\ref{sec:dynam}\\
DyRuRoHom & Dynamic Run-Time Robust Homogeneous& Sec.~\ref{sec:dynam}\\
StScHet & Heterogeneous with Strong-Scaling ($S>0$) & Sec.~\ref{sec:heter}\\
noStScHet & Heterogeneous without Strong-Scaling ($S=0$) & Sec.~\ref{sec:heter}\\\hline
\end{tabular}
\end{table}

\section{Scheduling comparison} \label{sec:sched}
%
In this section, we evaluate the sampling strategies from the previous section
and illustrate the influence of the
serial fraction parameter $B$, of the level-averaged number of sequential steps
and of the run-time variation.

\subsection{The influence of  the number of sequential steps}
The fact that
processor and sample numbers have to be integer not only complicates
the solution of the optimization problem, it also strongly influences
the amount of imbalance.

Let us start with some preliminary considerations and assume that
there are no run-time variations.
Now, let  $\Delta t \ge 0$ denote the relative
difference $\Delta t$ between the run-time $\sum_{\ell=0}^L
k_\ell^{\text{seq}}(\theta_\ell) t_{\ell,\scaleparam_\ell }$ of the
presented homogeneous strategies and the theoretically optimal
run-time in \eqref{eq:optruntime} (with $\mathbb{E}(C_{\ell,0}) =
1$). Using the MLMC level efficiency we can quantify $\Delta t$ as
\begin{eqnarray*}
\Delta t & = & \frac{P\foot{max}}{P_0^{\min}} \frac{\sum_{\ell=0}^L  k_\ell^{\text{seq}}(\theta_\ell) t_{\ell,\scaleparam_\ell }}{\sum_{\ell=0}^L N_{\ell} 2^{3\ell} t_{\ell,0}} -1 
=  \frac{\sum_{\ell=0}^L  N_\ell 2^{3\ell}   t_{\ell,0 }
  (\eta_\ell (\scaleparam_\ell))^{-1}}
{\sum_{\ell=0}^L N_{\ell} 2^{3\ell} t_{\ell,0}} -1 
\end{eqnarray*}
For the special case that  $P\foot{max} /(2^{3\ell +\scaleparam}
P_0^{\min}) \in \mathbb{N}$, we can further bound  $\Delta t$ in
terms of $k\foot{seq}  := \sum_{\ell=0}^L N_{\ell} 2^{3\ell} 
 P_0^{\min} /P\foot{max}$.  We assume that $t_{0,0} \le
t_{0,\ell} \le t_{0,L}$, for all $\ell = 0,\ldots, L$, which is
typically the case. The ratio $t_{L,0}/t_{0,0}$ reflects the weak scalability 
of the solver. Peta-scale aware massively parallel codes have a factor 
close to one. Recall from \eqref{eq:matrixss} that for our solver
$t_{L,0}/t_{0,0} = 179/167 \approx 1.07$.  
Since $k_\ell^{\text{seq}}(\scaleparam_\ell) t_{\ell,\scaleparam_\ell}
\leq  k_\ell^{\text{seq}}(0) t_{\ell,0} $ and since $k_\ell^{\text{seq}}(0) \le N_{\ell} 2^{3\ell}
\frac{P_0^{\min}}{P\foot{max}} + 1$ we have
\begin{eqnarray*}
\Delta t 
& \leq &  \left(\frac{  \max_{0 \leq \ell \leq L}
t_{\ell,0} }{   \min_{0 \leq \ell \leq L} t_{\ell,0}} \right)
\frac{P\foot{max}}{P_0^{\min}}
\frac{L+1}{ \sum_{\ell=0}^L N_{\ell} 2^{3\ell}} \ = \
         \frac{t_{L,0}}{t_{0,0}} \, \frac{L+1}{k\foot{seq}}.
\end{eqnarray*}
  The larger $k\foot{seq}$, the smaller the efficiency loss. 

\begin{figure}[ht]
\centering
  \includegraphics[clip=true, trim= 0mm 10mm 0mm 0mm,width=0.7\textwidth]{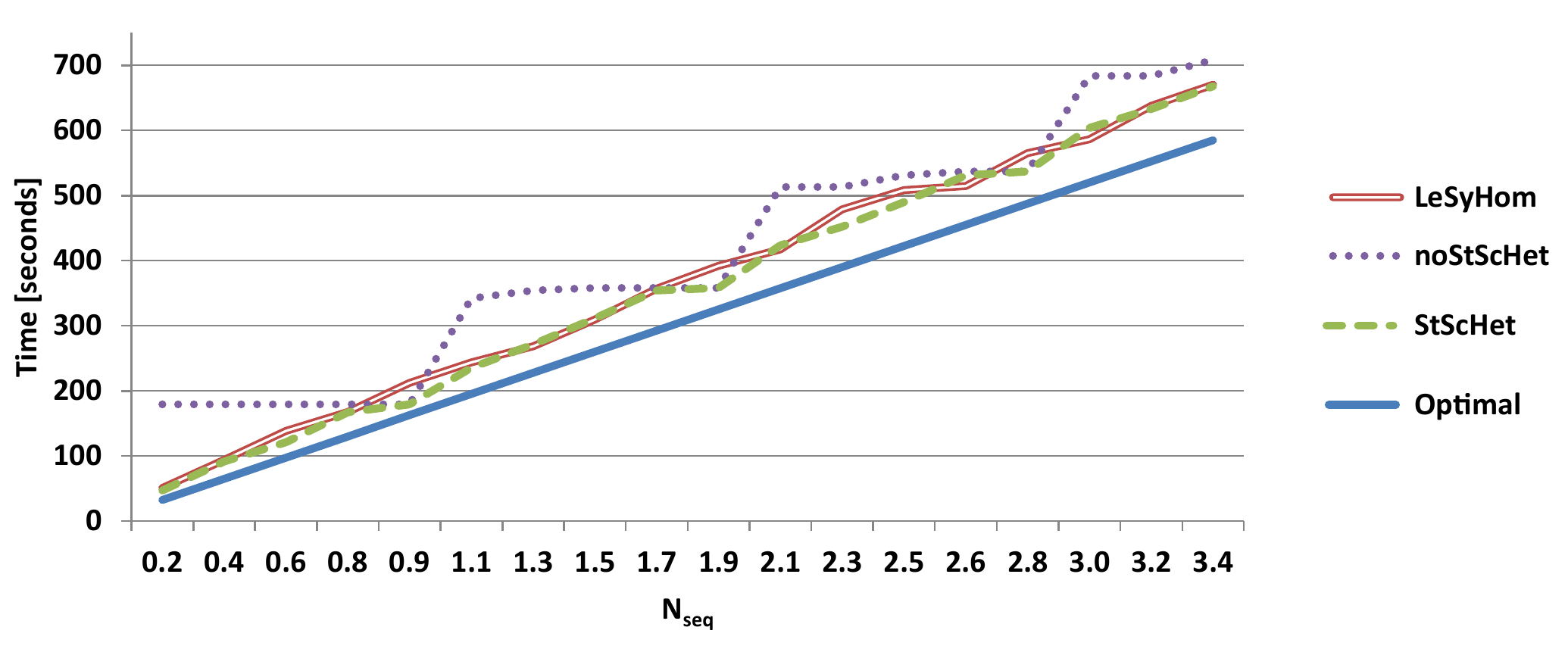}
\caption{\label{fig:nlvar}Heterogeneous versus homogeneous scheduling for $k\foot{seq} \in [0.2, 3.4]$.}
\end{figure}

In
Fig.~\ref{fig:nlvar} we compare  LeSyHom, noStScHet, StScHet and increase
$k\foot{seq} $ from $0.2$ to $3.4$.
All strategies stay within the theoretically predicted upper bound.
The two scheduling strategies, LeSyHom and StScHet, that exploit the
scaling properties of the solver are
significantly more robust with respect to $k\foot{seq} $ than the heterogeneous strategy, noStScHet,
for which we ignore the scalability window and set $S=0$.
This observation is particularly relevant for adaptive MLMC
strategies where $N_\ell$ may be increased within any of the
adaptive steps and then a new optimal scheduling pattern has to be 
identified.
For noStScHet, we observe a staircase pattern that 
is a direct consequence of the ceil operator. 
This  effect can be easily counterbalanced by exploiting the scalability window of the solver.
Moreover the run-times for LeSyHom and StScHet are larger than the
optimal one by roughly a factor of $1.5$ for $k\foot{seq} = 0.2$, 
but only by a factor of $1.15$ for $k\foot{seq} = 3.4$. 
Thus both these strategies are robust and efficient with respect to 
variations in $k\foot{seq}$.

 \subsection{The influence of solver scalability}
\begin{figure}[ht]
\centering
\resizebox{0.7\textwidth}{!}{\relsize{3} 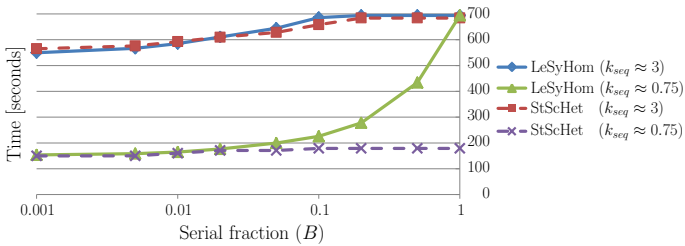}
\caption{\label{fig:strongvar}  Influence of the serial fraction parameter on the run-time. }
\end{figure}

The serial fraction parameter $B$ models the strong scaling 
of the solver, see Sec.\ \ref{sec:impl}.
The higher $B$, the less beneficial it is to increase $\scaleparam$.
In Fig.~\ref{fig:strongvar}, we consider the influence of $B$ on the run-time for two
different values of $k\foot{seq}$, namely $0.75$ and $3$, and compare LeSyHom and StScHet.

First, we consider the case  $k\foot{seq} \approx 0.75$.
With a serial fraction parameter $B \leq 0.02$, there is almost no run-time difference between the two strategies. For $B$ up to $0.1$ the run-time difference is below  $25\%$. 
However, for larger $B$, the run-time increases significantly for LeSyHom.
This can be explained by the fact that for large $B$, the strong scalability
property of the solver is too poor to obtain a robust scheduling
pattern, and only a heterogeneous strategy with its flexibility to schedule
in parallel samples on different levels can guarantee small run-times.
Homogeneous strategies provide enough flexibility to be efficient in
the case of large scalability windows with small values of $B$.
For a small value of $k\foot{seq}$,
the run-time of StScHet depends only very moderately on the serial 
fraction parameter $B$.

The situation is different for larger values of $k\foot{seq}$.
Then both strategies exhibit roughly the same performance,
but the total run-time is more sensitive to the size of $B$.
A good strong scaling of the PDE solver can improve the time to
solution by up to $27\%$ for the homogeneous and up to $21\%$ 
for the heterogeneous bulk synchronous case.
As expected, carrying out one synchronization step with
$k\foot{seq} \approx 3$ is more efficient than four steps 
with $k\foot{seq} \approx 0.75$. This observation is
important for the design of efficient adaptive strategies, i.e., they should
not be too fine granular.
For highly performant multigrid solvers, i.e., $B \leq 0.05$, the much
simpler  homogeneous strategies are an excellent choice, in particular
for $k\foot{seq} \geq 1$. On the other hand, when the
parallel performance of the solver is poorer, which is typically the case in the 
peta-scale regime, i.e. near the strong scaling limit of the multigrid
solver, the more complex heterogeneous strategies lead to
significantly better efficiency gains.

\subsection{Robustness and efficiency with respect to the parameters}

In this subsection, we modify all three key parameters that we have
discussed so far. We assume again that the run-time variations $C_{\ell,\scaleparam}(\cdot)$
are independent of $\ell$ and $\scaleparam$ and use a half-normal 
distribution to model $C_{0,0}(\cdot)$. More
specifically, we assume that $C_{0,0}(\cdot)-1$ follows a
half-normal distribution with parameter $\textit{Var}$, i.e.~its mode
is at~1. The time $t_{\ell,\scaleparam}$ is chosen to be the run-time of the mode,
as described in Sec.\ \ref{sec:parameters}. 

 \begin{figure}[ht]
\centering
\resizebox{0.9\textwidth}{!}{\relsize{3} 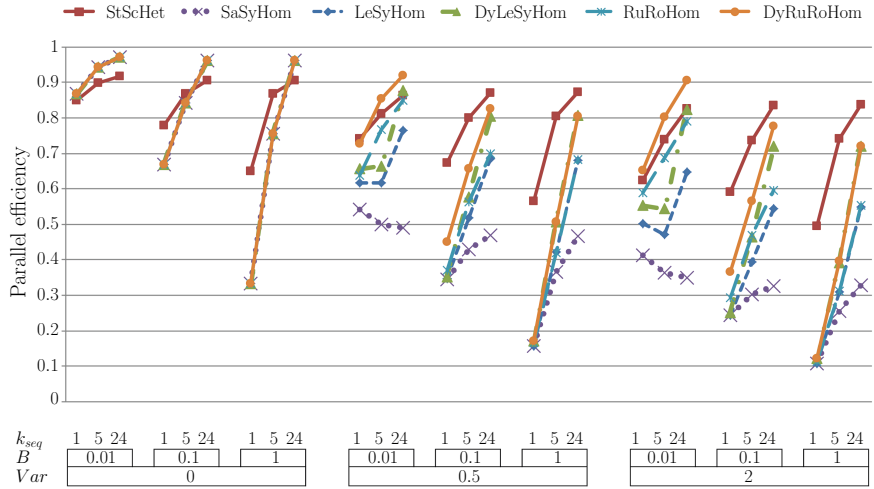}
\caption{\label{fig:compsched}MLMC efficiencies for different values
  of the parameters $k\foot{seq}$, $B$ and
$\textit{Var}$, and for all the different scheduling strategies.}
 \end{figure}
Fig.~\ref{fig:compsched} illustrates the parallel efficiencies of all the 
strategies developed above (cf.\ Tab.\ \ref{tab:summary}), as well as their
robustness with respect to the parameters $k\foot{seq}$, $B$ and
$\textit{Var}$,.
The parallel efficiency is calculated with respect to the
theoretical, optimal run-times given in~\eqref{eq:optruntime} not with 
actual measured run-times.
We choose $B \in \{0.01, 0.1, 1\}$ and $\textit{Var} \in \{ 0,0.5,
2\}$, and set the sample numbers on the different levels to be
$(N_0,N_1,N_2,N_3) = k\foot{seq} (1366, 228, 36, 5)$,
with $k\foot{seq} \in  \{0.98,4.92,23.60\} $
(abbreviated by $ \{1,5,24\} $ in Fig.~\ref{fig:compsched}).

We comment first on the case of no run-time variations, i.e.,
$\!\textit{Var}\!=0$, where our numerical results confirm 
that all homogeneous strategies produce
the same performance.
For large numbers of sequential steps,
the homogeneous variants are superior to the heterogeneous ones.
This is mainly due to the
constraint \eqref{eq:constraint1} which forces us to consider
all levels in parallel. As mentioned above, this constraint is not
essential and dropping it might lead to more efficient heterogeneous 
strategies. 
This will be the subject of future work.
If variations in the run-time are included,
then all homogeneous strategies yield different results.
The parallel efficiency of the simplest one, SaSyHom, then drops to 
somewhere between $0.1$ and $0.55$.
As expected, the worst performance is observed for
a small $k\foot{seq}$, poor solver scalability, and high run-time variation.
In that case, the dynamic variants can counterbalance the run-time
variations more readily and provide
computationally inexpensive 
scheduling schemes (provided the technical realization is 
feasible).

Secondly, we discuss the case of a small value of $k\foot{seq}$. This
typically occurs if the machine is large
or if the adaptive MLMC algorithm is used. Here,
only the heterogeneous strategies can guarantee acceptable
parallel efficiencies for all values of $B$. The homogeneous variants result 
in efficiencies below $0.7$ and $0.4$, for $\textit{Var}=0$ and for $B=0.1$ and  $B=1$,
respectively.

In all considered cases, one of our strategies results in parallel efficiencies of more than 
$0.5$; in many cases even more than $0.7$. 
For moderate run-time variations 
and large enough $k\foot{seq}$, the parallel efficiency of StScHet 
improves to more than $0.8$. StScHet is also
the most robust strategy with respect to solver scalability.
However, for solvers with good scalability, i.e.\ $B \leq 0.05$, 
the DyRoRuHom strategy is an attractive alternative, since it
does not require any sophisticated meta-heuristic scheduling algorithm
and can dynamically adapt to run-time variations in the samples.

\section{Numerical results for MLMC} \label{sec:error}
%
In this section, the scheduling strategies developed above are
employed in a large-scale MLMC computation. We
consider the model problem in Sec.~3 with $D = (0,1)^3$ and 
$f \equiv 1$, discretised by piecewise linear FEs. 
For the relevant problem sizes, the serial fraction parameter for 
our multigrid PDE solver is  $B \leq 0.02$, and the fluctuations in run-time are
$< 2\% $. Only few timings deviate substantially from the average
(c.f.~Fig.~\ref{fig:strongam}) so that we focus on investigating strategies for that regime.

The following experiments
were carried out on the peta-scale supercomputer
JUQUEEN, 
a 28 rack BlueGene/Q system located in J\"ulich, Germany%
\footnote{http://www.fz-juelich.de/ias/jsc/EN/Expertise/Supercomputers/JUQUEEN/JUQUEEN\_node.html}.
Each of the 28\,672 nodes has 16 GB main memory 
and 16 cores operating at a clock rate of 1.6 GHz. 
The compute nodes are connected via a five-dimensional torus network. 
HHG is compiled by the IBM XL C/C++ Blue Gene/Q, V12.0 compiler suite
with MPICH2 that implements the MPI-2 standard and supports RDMA. 
Four hardware threads can be used on each core to hide latencies. We 
always use 2 processes (threads) per core to maximize the 
execution efficiency.
\subsection{Static scheduling for scenarios with small run-time variations}
\label{sec:testcase1homogenbulk}
We choose four MLMC levels, i.e., $L=3$,
with
a fine grid that has roughly $1.1 \cdot 10^9$ mesh nodes. The random coefficient is
assumed to be lognormal with exponential covariance, $\sigma^2 = 1$
and $\lambda = 0.02$. The quantity of interest is the PDE solution $u$
evaluated at the point $x=(0.25, 0.25, 0.25)$.
All samples are computed using a fixed multigrid cycle structure with one
FMG-2V(4,4) cycle, i.e.,
a full multigrid method (nested iteration)
with two V-cycles per new level, as well as four pre- and 
four post-smoothing steps.
In \cite{gmeiner2015towards} it is shown that this multigrid method
delivers the solution of a scalar PDE with excellent numerical and parallel efficiency.
In particular, the example is designed such that
after completing the FMG-2V(4,4) cycle for all samples,
the minimal and maximal residual differ
at most by a factor $1.5$ within each MLMC level.
For the MLMC estimator, an a priori strategy is assumed, based on
pre-computed variance estimates, such that 
$(N_\ell)_{\ell=1,2,3,4} = (4\,123, 688, 108, 16)$. 
We first 
study the balance between sample and solver parallelism
and thus the tradeoffs between the efficiency of the parallel solver and 
possible load imbalances in the sampling strategy, as introduced
in Sec.\ \ref{sec:parameters}.
We set $P_0\head{min}=1$, and consequently
$P_{\ell}\head{min} = 2^{3\ell}$.
The run-times to compute a single sample with $P_{\ell}\head{min}$ processors
are measured as $(t_{\ell,0})_{\ell=0,1,2,3} = (166, 168, 174, 177)$
seconds, showing only a moderate increase in runtime and confirming the
excellent performance of the multigrid solver.

A lower bound for the run-time of the parallel MLMC estimator of
$t\head{opt}\foot{mlmc}=520\,\si{\second}$  is now provided by 
eq.\ \eqref{eq:optruntime}.  
A static cost model is justified since the timings 
between individual samples vary little.
We therefore employ the level synchronous homogeneous (LeSyHom)
scheduling strategy, as introduced in Sect.~\ref{sec:SaSyHom}. 
This requires only a few, cheap real time measurements
to configure the MLMC scheduling strategy. 
The smaller $\scaleparam$, the larger the solver efficiency $\text{Eff}_\ell(\scaleparam)$while the larger $\scaleparam $, the smaller $\text{Imb}_\ell(\scaleparam)$.
To study this effect quantitatively, we measure the parallel solver
efficiency. Here, we do not use \eqref{eq:amdahls}, but actual
measured values for $t_{\ell,\scaleparam}$ instead, and we find that
on level $\ell=0$ we have
$\{\text{Eff}_{0} (0), \text{Eff}_0(1), \text{Eff}_0(2), \text{Eff}_0(3), \text{Eff}_0(4) \}  = \{1, 0.99, 0.96, 0.92, 0.86\}$. 
%

In 
Tab.~\ref{tab:loadeff}, we present for each level the efficiency
$\eta_\ell(\scaleparam)$, see \eqref{eq:effs}, and the run-time as a
function of $\scaleparam$. For $\scaleparam = 0$ and $\ell=0$, the
runs are carried out with one thread, and they go up to  8\,192 hardware
threads on 4\,096 cores for $\scaleparam =4$ and $\ell=3$.
We see a very good correlation between
predicted efficiencies and actual measured times in Tab.~\ref{tab:loadeff}. 
The maximal efficiency and the minimal run-time on each level are
marked in boldface to highlight the best setting.
\begin{table}[th]
\caption{\label{tab:loadeff}
Level and total run-time and efficiency of MLMC for fixed $\scaleparam$.}
\centering
\begin{tabular}{l|rr|rr|rr|rr|rr}
\hline
$\scaleparam$ & time & $\eta_0 (\scaleparam)$ 
& time & $\eta_1 (\scaleparam)$
& time & $\eta_{2} (\scaleparam)$
& time & $\eta_{3} (\scaleparam)$ & time & $\eta(\scaleparam)$\\
 \hline
0 & 167  & 0.50    & 171   & 0.67    & 177   & 0.84    & \textbf{179} & \textbf{1.00}  &694  & 0.75    \\
1 & 168  & 0.50    & 173  & 0.67    & 181  & 0.84    & 183 & 0.99  &  704& 0.74    \\
2 & 127  & 0.64    & \textbf{134}  & \textbf{0.86}    & 188  & 0.81    & 193& 0.96  & 642 & 0.81    \\
3 & 108  & 0.74    & 139  & 0.83    & \textbf{169}  & \textbf{0.89}    &  199& 0.92   & \textbf{615}& \textbf{0.85}    \\
4 & \textbf{104} & \textbf{0.77}    & 136 & 0.84    & 181 & 0.81  &  
218& 0.86   & 640 & 0.83   \\ \hline
\end{tabular}
\end{table}

Keeping $\scaleparam$ fixed, the minimal run-time is
$615$. Our homogeneous scheduling strategies pick $\scaleparam_\ell$ for each
level automatically and by doing so, a considerably shorter
run-time of $586 $ is obtained, increasing the efficiency to
$0.89$. 
If it were possible to scale the solver perfectly also to processor
numbers that are not necessarily powers of 2, we could
reduce the compute time even further 
by about $11\%$ from $586$ to $520$ seconds.
In summary, we see that exploiting the strong scaling
of the PDE solver helps to avoid load imbalances 
due to oversampling and improves the time to solution by about $15\%$
from $694$ to $586$ seconds. 
The cost is well distributed across all
levels, although most of the work is on the finest level which is
typical for this model problem (cf.~\cite{CST:2011}).  

\begin{figure}
\centering
  \includegraphics[trim= 0mm 0mm 0mm 0mm,width=0.6\textwidth]{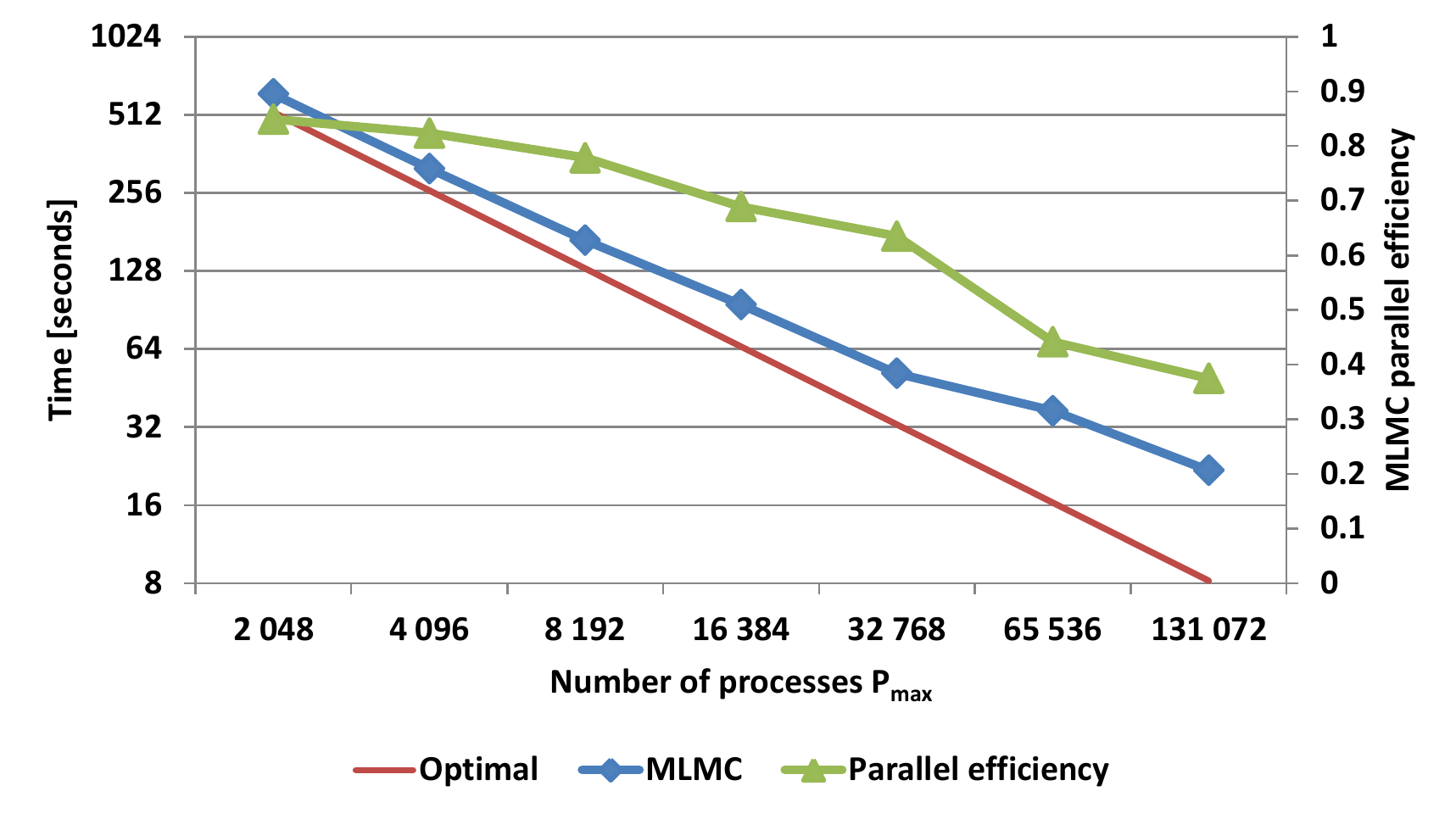}
\caption{\label{fig:mlmcstrong}Strong scaling of MLMC using homogeneous bulk synchronous scheduling.}
\end{figure}

We conclude this subsection with a strong scaling experiment,
i.e., we increase the number of processes in order to reduce the overall time to solution.
Since we are interested to analyse the behavior for extremely large
$P\foot{max} $, we reduce the number of samples to 
$N_l = (1\,031, 172, 27, 4)$. The time for one
MLMC computation with an increasing number of processes $P\foot{max}$
is presented in Fig.~\ref{fig:mlmcstrong}.
The initial computation employs $P\foot{max} = 2\,048$
which is large enough so that all fine-grid samples can be computed concurrently.
We scale the problem up to 131\,072 processes.
Increasing $P\foot{max}$ while keepimg the number of samples per level
fixed results in a decrease of $k\foot{seq}$.
Thus the load imbalance increases and the total parallel efficiency
decreases. Nevertheles even with $P\foot{max} =32768$,
we obtain a parallel efficiency over $60\%$  while for $P\foot{max} =
131072$ the eficiency drops below $40\%$. This can be circumvented by
an increase of the size $S$ of the sclability window. 
 Overall the compute time for the MLMC estimator can be 
reduced from $616$ to $22$ seconds. 
For each choice of $P\foot{max}$, we select the optimal regime for
$\theta_\ell$, $\ell = 0,\ldots,3$, as discussed above. 
Together with the excellent strong scaling behavior of the parallel HHG 
multigrid solver this leads to the here demonstrated combined parallel
efficiency of the MLMC implementation.

\subsection{Adaptive MLMC}
Finally, we consider an adaptive MLMC algorithm as introduced in Sec.\ \ref{sec:adaptive} 
in a weak scaling scenario,
i.e., increasing the problem size proportionally to the processor count. 
%
\begin{table}[ht]
\centering%
\caption{\label{fig:adaptivemlmc} Weak scaling of an adaptive MLMC estimator.}%
\begin{tabular}{rrrrrrr}
\hline
            &    & &  \multicolumn{2}{c}{No. Samples}   & Correlation  & Idle  \\
Processes & Resolution & Runtime & Fine  & Total &  length &  time \\ \hline
4\,096   & $1\,024^3$      & $5.0 \cdot 10^3$ s        & 68             & 13 316           & 1.50E-02           & 3\%                \\

32\,768 & $2\,048^3$      & $3.9 \cdot 10^3$  s      & 44             & 10 892           & 7.50E-03           & 4\%                \\
262\,144  & $4\,096^3$      & $5.2 \cdot 10^3$   s     & 60             & 10 940           & 3.75E-03           & 5\%        \\      
\hline 
\end{tabular}
\caption{Number of samples and over-samples for different levels for the largest run.}
\label{tab:oversampling}
\begin{tabular}{rrrrrrr}
\hline
          &    &  \multicolumn{2}{c}{No. Samples}   &   \multicolumn{2}{c}{No. Over-samples}  \\
Level & No. partitions & Scheduled & Calculated & Estimated & Actual \\ \hline
0 & 2 048 & 7 506 & 8 192 & 3 726 & 686 \\
1 & 256  & 2 111 & 2 304 & 429  & 193 \\
2 & 32   & 382  & 384  & 15   & 2   \\
3 & 4    & 57   & 60   & 3    & 3          
\\
\hline
\end{tabular}
\end{table}
Each row in Tab.\ \ref{fig:adaptivemlmc} summarizes one adaptive MLMC computation. 
The MLMC method is initially executed on 2\;048 cores and $P_{\ell}$ is chosen as 
$P_{\ell=1,2,3,4} = (2, 16, 128, 1\,024)$, 
In each successive row of the table, the 
number of unknowns on the finest level in MLMC and the number of
processors on each level is increased by a factor of eight. 
Moreover, the correlation length $\lambda$ of the coefficient field is reduced by a factor
of two ($\sigma^2= 1$ is kept fixed). 
This means that the problems are actually getting more
difficult as well.
The quantity of interest is defined as
the flux across a separating plane $\Gamma$ at $x_2=0.25$, i.e.,
$$
Q(u,\omega) = \int_\Gamma k(x,\omega) \frac{\partial u}{\partial n} ds\,.
$$
In all cases, the initial number of samples is set to 
$N_{\ell=1,2,3,4} = (1\,024, 256, 64, 16)$. The final
number of samples is then chosen adaptively by the MLMC algorithm. It  is 
listed in the table. As motivated in Sec.\ \ref{sec:mlmc}, the
tolerance for the sampling
error, which is needed in \eqref{quasioptimal_Nl} to adaptively
estimate $N_\ell$, is chosen as 
\mbox{$\varepsilon_s \approx |\mathbb{E}[Q_{L}-Q_{L-1}]|$}, balancing
the sampling error with the bias error.
\begin{figure}[th]
\centering
\resizebox{0.40\textwidth}{3.8cm}{\relsize{6} 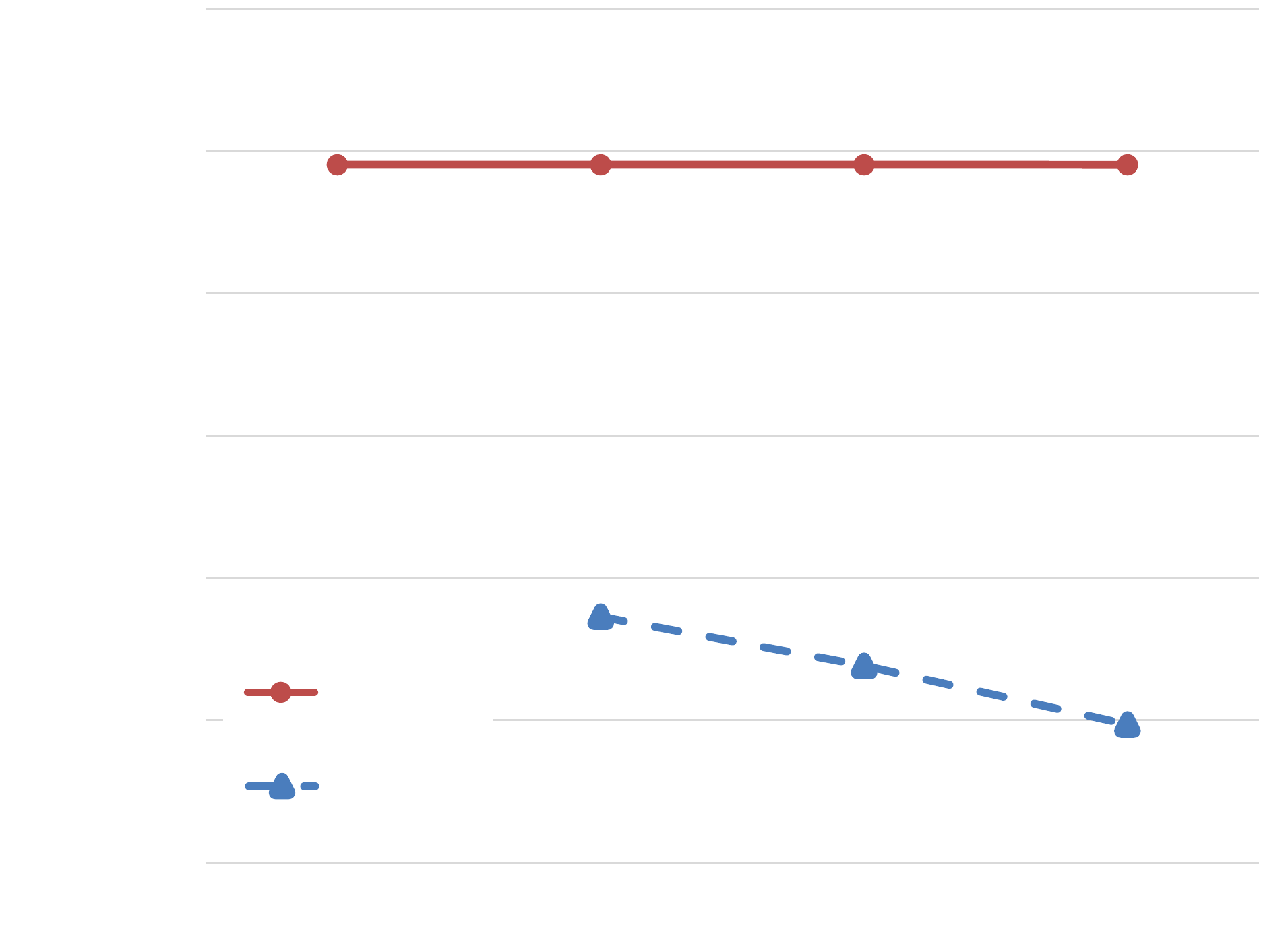}
\hspace*{1cm}
\resizebox{0.40\textwidth}{3.8cm}{\relsize{6} 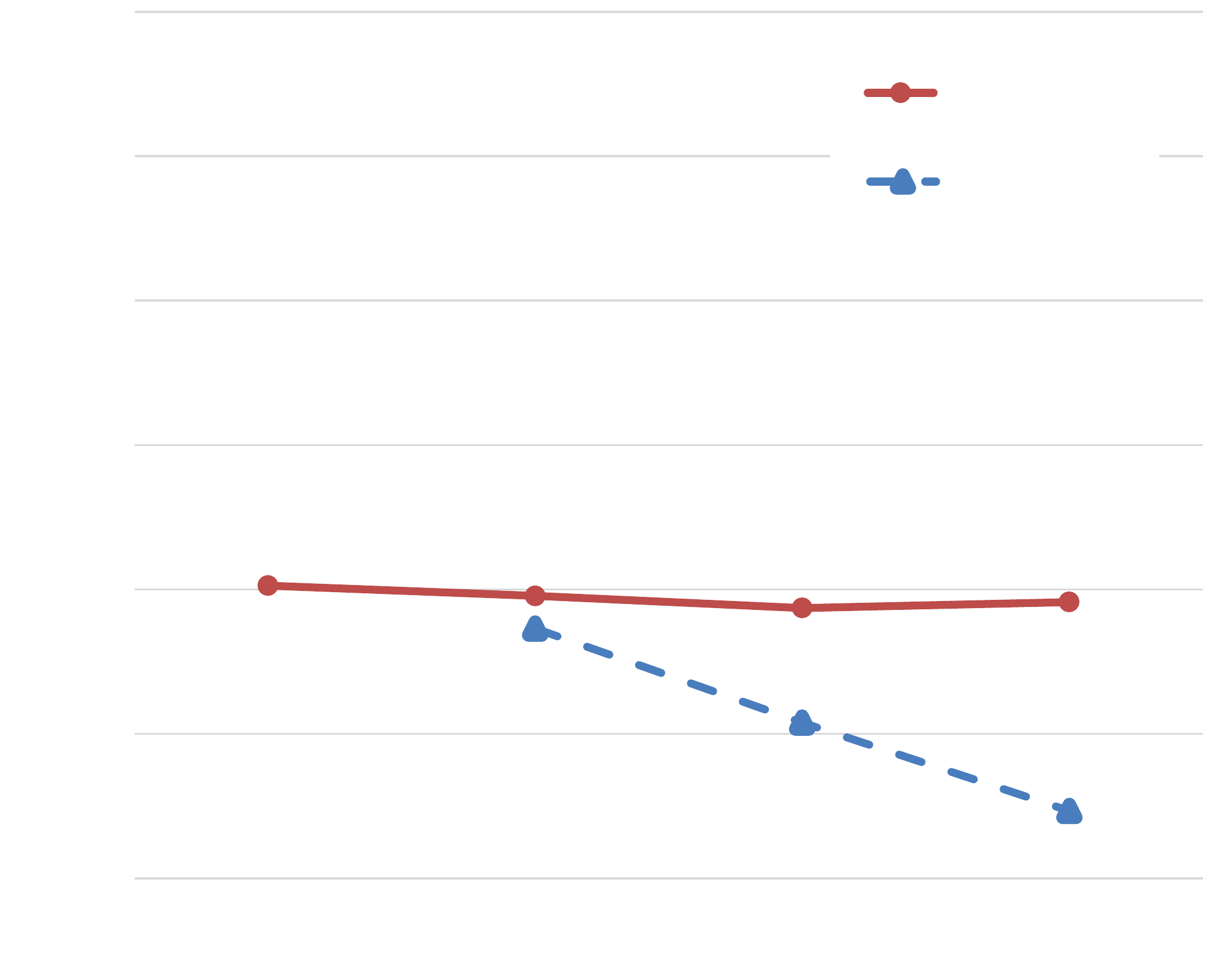}
\caption{\label{fig:adaptmlmcexpvar} MLMC performance plots: expected value (left) and
  variance (right) of $Q_\ell$ (red, solid) and $Y_\ell$ (blue,
  dashed) for
$\lambda=0.015$ and $\sigma^2= 1$.}
\end{figure}
The estimates for the expected values and for the variances of
$Q_\ell$ and $Y_\ell$, for a problem of size $M_\ell = 1\,024^3$
and with a correlation length of $\lambda = 0.015$, are plotted in 
Fig.~\ref{fig:adaptmlmcexpvar}. The expected values and
the variances of $Y_\ell$ show the expected asymptotic behavior as
$\ell$ increases, confirming the benefits of the multilevel approach.
The total number of samples that are computed is 13\,316, but only
68 of them on the finest grid. A standard Monte Carlo estimator would
require several thousand samples on level 3 and would be significantly
more costly. The idle time, in the last column of 
Tab.\ \ref{fig:adaptivemlmc},
accounts for the variation in the number of V-cycles, 
required to achieve
a residual reduction of $10^{-5}$ on each level within each call to
the FMG multigrid algorithm. 

The largest adaptive MLMC
computation shown in Tab.\ \ref{fig:adaptivemlmc} involves a finest grid
with almost \num{7e10} unknowns. Discrete systems of this size 
must be solved $60$ times, together with more than 10\;000 smaller 
problems, the smallest of which still has more than \num{1.6e7} 
unknowns.
With the methods developed here, a computation of such 
magnitude requires a compute time of less than 1.5 hours when 
131\;072 cores running 262\,144 processes are employed.
Additional details for this largest MLMC computation are presented in
Tab.~\ref{tab:oversampling}. The
table lists the number of partitions that are used on each level for
the respective problem sizes. The number of calculated samples on each 
level is a multiple of the number of these partitions.
As Tab.\ \ref{tab:oversampling} illustrates, the number of scheduled 
samples is smaller and the difference indicates the amount of oversampling.
The number of unnecessary samples is presented explicitly in the last
column of the table to compare with the estimated number of unneeded samples. 
This estimated number is significantly higher on each level, since it
is the sum of all the oversampled computations in all stages of the
adaptive MLMC algorithm. 
As we pointed out earlier, this is caused by a special feature of the
adaptive MLMC algorithm. Samples that were predicted to be redundant
in an early stage of the algorithm, may become necessary later in the computation.
Thus at termination, the actual oversampling is significantly less than predicted.
This is a dynamic effect that cannot be quantified easily in a static a priori fashion.

\section{Conclusions}
In this paper we have explored the use of multilevel Monte Carlo methods
on very large supercomputers.
Three levels of parallelism must be coordinated, since it is not sufficient
to just execute samples in parallel. The combination of solver- and
sample-parallelism leads to a non-trivial scheduling problem, where the trade-off
between solver scalability, oversampling, and 
additional efficiency losses due to run-time variations
must be balanced with care.
This motivated the development of scheduling strategies of increasing complexity,
including advanced dynamic methods that rely on meta-heuristic search algorithms. 
These scheduling algorithms are based on performance predictions
for the individual tasks that can in turn be derived from run-time
measurements and performance models motivated by Amdahl's law.

The success of the techniques and their scalability are demonstrated
on a large-scale model problem. The largest MLMC computation involves
more than 10\,000 samples and a fine grid resolution with almost 
\num{7e10} unknowns. It is executed
on 131\,072 cores of a peta-scale class supercomputer in 
1.5 hours of total compute time.

\bibliographystyle{abbrv}
\bibliography{literature}

\end{document}